\documentclass[aps, twocolumn,prl,showpacs,amsmath,amssymb,superscriptaddress]{revtex4-1}
\usepackage{amsmath,amsfonts,amssymb,graphics,graphicx,epsfig,color,times,indentfirst,layout,hyperref,subfigure,multirow,bm}
\usepackage{hyperref,soul}
\usepackage{ulem}
\hypersetup{linktocpage,,colorlinks=true}
\usepackage{threeparttable}

\usepackage{tikz}
\usepackage{tikz-cd}
\usetikzlibrary{arrows}
\usetikzlibrary{intersections}
\usetikzlibrary{shapes.geometric}
\usetikzlibrary{decorations.pathmorphing, patterns,shapes}
\usetikzlibrary{decorations.markings}

\def\beq{\begin{equation}}
\def\eeq{\end{equation}}
\def\bea{\begin{eqnarray}}
\def\eea{\end{eqnarray}}

\begin{document}
\title{Entanglement spectrum and entropy in topological non-Hermitian systems and non-unitary conformal field theory}

\author{Po-Yao Chang}
\affiliation{Department of Physics, National Tsing Hua University, Hsinchu 30013, Taiwan}
\affiliation{Max Planck Institute for the Physics of Complex Systems, N\"{o}thnitzer Strasse 38, D-01187 Dresden, Germany}
\affiliation{Department of Physics and Astronomy, Center for Materials Theory, Rutgers University, Piscataway, NJ 08854 USA}
\author{Jhih-Shih You}
\affiliation{Institute for Theoretical Solid State Physics, IFW Dresden, Helmholtzstr. 20, 01069 Dresden, Germany}
\author{Xueda Wen}
\affiliation{Department of Physics, Massachusetts Institute of Technology, Cambridge, MA 02139, USA}
\author{Shinsei Ryu}
\affiliation{James Franck Institute and Kadanoff Center for Theoretical Physics, University of Chicago, Illinois 60637, USA }

\begin{abstract}
  We study entanglement properties of free-fermion systems without
  hermiticity by use of
  correlation matrix and overlap matrix in the biorthogonal basis.
  We find at a critical point in the non-Hermitian Su-Schrieffer-Heeger (SSH) model with parity and
  time-reversal symmetry (PT-symmetry)
  the entanglement entropy exhibits 
  a logarithmic scaling with corresponding central charge $c=-2$,
  signaling the emergence of non-unitary conformal field theory.
  In addition,
  we demonstrate that,
  in the PT-symmetric SSH model and the non-Hermitian Chern insulators,
  the entanglement spectrum characterizes the topological properties
  in terms of the existence of mid-gap states.
\end{abstract}

\maketitle

\section{Introduction}

In contemporary condensed matter physics, quantum entanglement plays a pivotal role
in characterizing and obtaining a deeper understanding of many-body quantum systems.
For example, the topological entanglement entropy provides a direct method
to detect topological order
\cite{Kitaev2006,Levin2006}.
The subsystem size scaling of the entanglement entropy
can be used to extract useful information of conformally invariant systems
\cite{Calabrese_2004}.
Ground-state properties of topological systems are also reflected in their entanglement spectra
\cite{2006PhRvB..73x5115R, Li2008, 2010PhRvB..81f4439P, Turner2010,Hughes2011,Chang_2014}.
Moreover, quantum entanglement paves a way for a deeper understanding 
of the renormalization group and an emergent 
holographic space-time 
\cite{Ryu2006, Ryu2006_2,Nishioka2009}.

In this letter, we aim to extend the quantum-entanglement based approach
to systems that lack hermiticity.
Non-Hermitian quantum mechanical systems, in particular, those which host topological phenomena, 
have attracted a lot of interest recently
\cite{Esaki2011,Schomerus2013,Lee2016, Weimann2017,Alvarez2018, Yin2018, Kunst2018, Kawabata2018_2,Klett2017,Klett2018,Yao2018,Lieu2018,Gong2018, Kawabata2019,Wang2019,Parto2018,Shen2018,Yoshida2018,Yoshida2019}.
To be concrete,
we take, as paradigmatic examples, 
the non-Hermitian Su-Schrieffer-Heeger (SSH) model with the combination of parity and
time-reversal symmetry (PT-symmetry)
\cite{Klett2017, Klett2018,Yao2018,Lieu2018},
and non-Hermitian Chern insulators
\cite{Kawabata2018}.
These non-Hermitian models host a variety of trivial and topological gapped phases,
as well as gapless phases and critical points.

In order to introduce and investigate the entanglement entropy/spectrum
in these non-Hermitian systems,
we use a generalized definition of the (reduced) density matrix in terms
of the biorthogonal basis
\cite{Brody2013, Weigert2003}.
First, we will show that for non-interacting systems, 
the entanglement entropy and spectrum 
can be efficiently computed from the
correlation and overlap matrices in the biorthogonal basis.
Furthermore, we will demonstrate that the entanglement entropy and spectrum
can be used to detect various universal properties of
non-Hermitian gapped (topological) systems.
For example,
we show that the topological gapped (trivial) PT-symmetric phase
in the non-Hermitian SSH model supports (does not support)
robust mid-gap states in the entanglement spectrum.
The existence of the mid-gap states in the entanglement spectrum is concurrent 
with the existence of protected physical boundary modes in these PT-symmetric phases
\cite{Yao2018,Lieu2018}:
Thus, the entanglement spectrum provides a way to characterize the topological non-Hermitian phases.

Furthermore, we found, at a critical point appearing in the non-Hermitian SSH model,
the entanglement entropy {\it decreases} logarithmically in the subsystem size --
this signals the emergence of non-unitary conformal field theory (CFT).
%
%
More specifically,
at the critical point separating the trivial PT-symmetric phase
and the spontaneously PT-broken phase in the non-Hermitian SSH model, 
we show that the entanglement entropy for the subsystem of length
$L_A$
scales as 
$S_A = (c/3) \ln L_A +\cdots$
with 
$c=-2$.
The negative central charge $c=-2$ can be attributed to
the {\it bc}-fermionic ghost theory
\cite{Polchinski,Blumenhagen,FRIEDAN198693, KAUSCH2000513, Kausch1995,GURUSWAMY1998661,SM}.

The entanglement entropy in non-unitary CFTs has been studied previously
\cite{Narayan2016,Bianchini2015,Bianchini2015a,Dupic2018,
  Bianchini_2016,Jatkar2017,Couvreur2017},
but in most cases, the entanglement entropy scaling is captured by 
the effective central charge, $c_{{\rm eff}}>0$,
even when the central charge itself is negative. 
I.e., the entanglement entropy in these cases increases logarithmically in the subsystem size,
$\sim (c_{{\rm eff}}/3)\ln L_A$.
In Ref.\ \cite{Couvreur2017},
the authors studied loop models by using the biorthogonal reduced density matrix
to compute the entanglement entropy, and found that
the entanglement entropy scaling is given by the central charge $\sim (c/3)\ln L_A$ with $c<0$.
In this letter, we will show that
the negative central charge can also occur
in non-Hermitial topological PT-symmetric systems \cite{Klett2017,Klett2018, Weimann2017, Kawabata2018_2, Yao2018,Lieu2018, Wang2019}.
Our findings shed new light on the nature of critical points appearing in non-Hermitian systems,
and in particular those in proximity to topological non-Hermitian phases.

We also found, at the other critical point of the PT-symmetric SSH model, 
i.e.,
the one which separates the topological PT-symmetric phase and
the spontaneously PT-broken phase, the entanglement entropy/spectrum
is somewhat more complicated and interesting;
There are two additional mid-gap states in the entanglement spectrum which mimic the physical boundary modes at this critical point. 
Nevertheless, 
by modifying the bipartition, we show that the entanglement scaling 
again is given by the logarithmic law with $c=-2$.
This reminds us of the similar behavior at the symmetry-enriched critical point found
in a Hermitian system
\cite{Verresen2019}.

We also demonstrate the quantum entanglement 
in the spontaneously PT-broken phase can also be studied,
once we consider a proper "ground state" which is
obtained by filling modes with real energy.
The entanglement entropy scaling gives 
the central charge $c=1$. 
We show the Jordan block form at the exceptional point leads to 
the ground state identical to the free Dirac theory.
Finally, as yet another example,
we study the entanglement spectrum of the non-Hermitian Chern insulators.
The mid-gap states in the entanglement Hamiltonian mimic the physical boundary modes 
in non-Hermitian systems.
Our method provides an alternative way to study the entanglement properties in both critical and topological non-Hermitian systems.



\section{Biorthogonal basis and non-Hermitian Hamiltonian}


Before discussing quantum entanglement of specific non-Hermitian lattice systems,
we start by developing a convenient method to
compute the entanglement spectrum/entropy in the biorthogonal basis,
which is valid for free (quadratic) systems.
Specifically,
we will show that the generalized reduced density matrix can solely be constructed
from the correlation matrix or the overlap matrix,
much the same way in regular Hermitian systems.
Our method is complementary to the flatten singular-value decomposition recently proposed in \cite{Herviou2019}.

Consider a generic quadratic non-Hermitian Hamiltonian,
$H = \sum_{i j} \phi^\dagger_{i} \mathcal{H}^{\ }_{i j} \phi^{\ }_{j}$,
with $\mathcal{H} \neq \mathcal{H}^\dagger$ and $\{ \phi^{\ }_{i}, \phi^\dagger_{j} \} =\delta_{ij}$ 
 being fermionic operators.
The biorthogonal basis is constructed from the 
left and right eigenvectors 
of ${\cal H}$, 
$\mathcal{H} |R_\alpha \rangle = \epsilon_\alpha |R_\alpha \rangle$, 
$\mathcal{H}^\dagger |L_\alpha \rangle = \epsilon_\alpha ^* |L_\alpha \rangle$, 
such that $\langle L_\alpha  | R_\beta \rangle = \delta_{\alpha \beta}$ \cite{Brody2013}.
The single-particle Hamiltonian can then be written as
$\mathcal{H}= \sum_\alpha \epsilon_\alpha |R_\alpha \rangle \langle L_\alpha |$,
and, correspondingly, the Hamiltonian can be written as 
\begin{align}
H &= \sum_\alpha \epsilon_\alpha \sum_i (R_{\alpha i} \phi_i)^\dagger \sum_j (L_{\alpha j} \phi_j) = \sum_\alpha \epsilon_\alpha \psi_{R\alpha}^\dagger \psi^{\ }_{L\alpha}, 
\label{Eq1}
\end{align}
where $\psi_{R\alpha}^\dagger$ and $\psi_{L\alpha}^\dagger$ are the right and left creation operators such that
$|R_\alpha \rangle = \psi_{R\alpha}^\dagger |0\rangle$,
$|L_\alpha \rangle = \psi_{L\alpha}^\dagger |0\rangle$, and  
$\psi_{R \alpha} |0\rangle= \psi_{L_\alpha} |0\rangle =0$.
It should be noted that $\psi_{L\alpha}$ and $\psi_{R\beta}$
are not ordinary fermionic operators, 
in the sense that they satisfy the 
commutation relationship $\{ \psi^{\ }_{L \alpha}, \psi^\dagger_{R \beta} \} =\delta_{\alpha\beta}$.
We refer these operators the bi-fermionic operators.



Many-body eigenstates can be constructed by acting on the vacuum with a set of ("occupied")
bi-fermionic creation opreators.
For example,  
$|G_R\rangle = \prod_{\alpha \in {\rm occ.}} \psi_{R \alpha}^\dagger | 0 \rangle$ 
satifies
$
H |G_R\rangle
= \sum_\alpha \epsilon_\alpha \psi_{R\alpha}^\dagger \psi^{\ }_{L\alpha} \prod_{\beta \in {\rm occ.}} \psi_{R\beta} ^\dagger |0\rangle = \sum_{\beta \in {\rm occ.}} \epsilon_\beta  |G_R\rangle.
$
Similarly, a many-body left eigenstate
$|G_L\rangle = \prod_{\alpha \in {\rm occ.}} \psi_{L\alpha}^\dagger |0\rangle$ 
satisfies 
$H^\dagger |G_L\rangle = \sum_{\alpha \in {\rm occ.}}\epsilon^*_\alpha |G_L\rangle$.


From many-body left and right eigenstates,
we can construct a non-Hermitian density matrix
as $\rho = |G_R\rangle \langle G_L|$ such that $\rho^\dagger \neq \rho$ and $\rho^2 =\rho$.
With this generalized notion of the density matrix,
we can introduce measures of quantum entanglement
by partitioning the total system into two subsystems $A$ and $B$,
and then by taking the partial trace over subsystem $B$, 
$\rho_A= {\rm Tr}_B\, \rho $.
We can discuss the spectrum of the generalized reduced density matrix, and the entanglement entropy.

In Hermitian systems, if the Hamiltonian has the quadratic form,
the reduced density matrix can be constructed either from the correlation matrix
\cite{Chung2001,Peschel2009}
or the overlap matrix
\cite{Klich2006}. 
We extended these derivations to non-Hermitian systems
with quadratic form in terms of bi-fermionic operators as follows\cite{SM}.

-- The correlation matrix is defined as 
\begin{align}
C_{ij} &= \langle G_L |  \phi_i^\dagger \phi_j  |G_R\rangle =\sum_{\alpha \in {\rm occ.}} L_{\alpha i} R^{\dagger}_{\alpha j}.
\end{align} 
The entanglement Hamiltonian 
${\cal H}^A$ can be introduced by
$C=e^{-\mathcal{H}^A}/(1+e^{-\mathcal{H}^A})$.
The entanglement entropy 
for subsystem $A$
can then be introduced as
$S_A 
 = -\sum_{\delta} \left[ \xi_\delta \ln \xi_\delta +(1-\xi_\delta) \ln (1-\xi_\delta)  \right]$,
where $\xi_\delta$ are the eigenvalues of $C_{ij}$.

-- The overlap matrix is defined as  
\begin{align}
M^A_{\alpha\beta} 
=\sum_{i \in A}  L_{\alpha i}^\dagger R^{\ }_{\beta i} 
= \sum_\delta p_\delta   (L^A_{ \alpha \delta} )^\dagger R^A_{\beta \delta}, 
\end{align}
 where $L^A_{ \alpha \delta}$ and $R^A_{ \beta \delta}$ are the corresponding left and right eigenvectors of $M^A$ with eigenvalues $p_\delta^*$ and $p_\delta$.
The original left and right 
many-body wavefucntions
as well as the reduced density 
matrix can be expressed 
in this new basis.
In particular, $\rho_A$
is given by
\begin{align}
\rho_A =  \bigotimes_\delta [p_\delta |L^A_\delta   \rangle \langle R^A_\delta | +(1-p_\delta) |0\rangle \langle 0 |].
\end{align}
The entanglement entropy can be directly obtained from the reduced density matrix as 
$S_A = - \sum_\delta [p_\delta \ln p_\delta + (1-p_\delta) \ln (1-p_\delta)]$.

%
%

\section{Entanglement entropy and entanglement spectrum in non-Hermitian systems}

\subsection{Non-Hermitian SSH model}


We now study the non-Hermitian SSH model
\cite{Lieu2018,Liang2013}
with the PT-symmetry defined in momentum space by
\begin{align}
\mathcal{H}_k =\left(\begin{array}{cc}i u & v_k \\v^*_k & -i u\end{array}\right),
\label{Eq:SSH}
\end{align}
where $v_k = w e^{-i k} + v$ with $u, v, w \in \mathbb{R}$ 
and $k$ is the single-particle momentum.
Here the PT-symmetry is defined as
$\sigma_x {\cal H}_k \sigma_x= {\cal H}_k^{\ast}$.
The energy is $E_k = \pm \sqrt{|v_k|^2 -u^2}$.
The right eigenvectors are
\begin{align}
|R_{k,+}\rangle = \left(\begin{array}{c}\frac{v_k}{|v_k|} \cos \frac{\phi_k}{2} \\ \sin \frac{\phi_k}{2}\end{array}\right),
\quad
|R_{k,-}\rangle = \left(\begin{array}{c} - \frac{v_k}{|v_k|} \sin \frac{\phi_k}{2} \\ \cos \frac{\phi_k}{2}\end{array}\right),
\end{align}
where $\mathcal{H}_k |R_{k,\pm} \rangle  = \pm E_k |R_{k _\pm} \rangle$ and $\phi_k = \tan^{-1} [{|v_k|}/{i u}]$,
and the corresponding left eigenvectors are
\begin{align}
|L_{k,+}\rangle = \left(\begin{array}{c}\frac{v_k}{|v_k|} \cos^* \frac{\phi_k}{2} \\ \sin^* \frac{\phi_k}{2}\end{array}\right),
\quad
|L_{k,-}\rangle = \left(\begin{array}{c} - \frac{v_k}{|v_k|} \sin^* \frac{\phi_k}{2} \\ \cos^* \frac{\phi_k}{2}\end{array}\right),
\label{Eq:L}
\end{align}
with $\mathcal{H}^\dagger_k |L_{k,\pm} \rangle  = \pm E_k |L_{k,\pm} \rangle$. 
The right and left eigenvectors satisfy
the biorthogonal condition, $\langle L_{k,\pm} | R_{k, \pm}\rangle =1$ and $\langle L_{k,\mp} | R_{k, \pm}\rangle =0$.



\begin{figure}[t]
  \centering
  \includegraphics[width=0.5\textwidth] {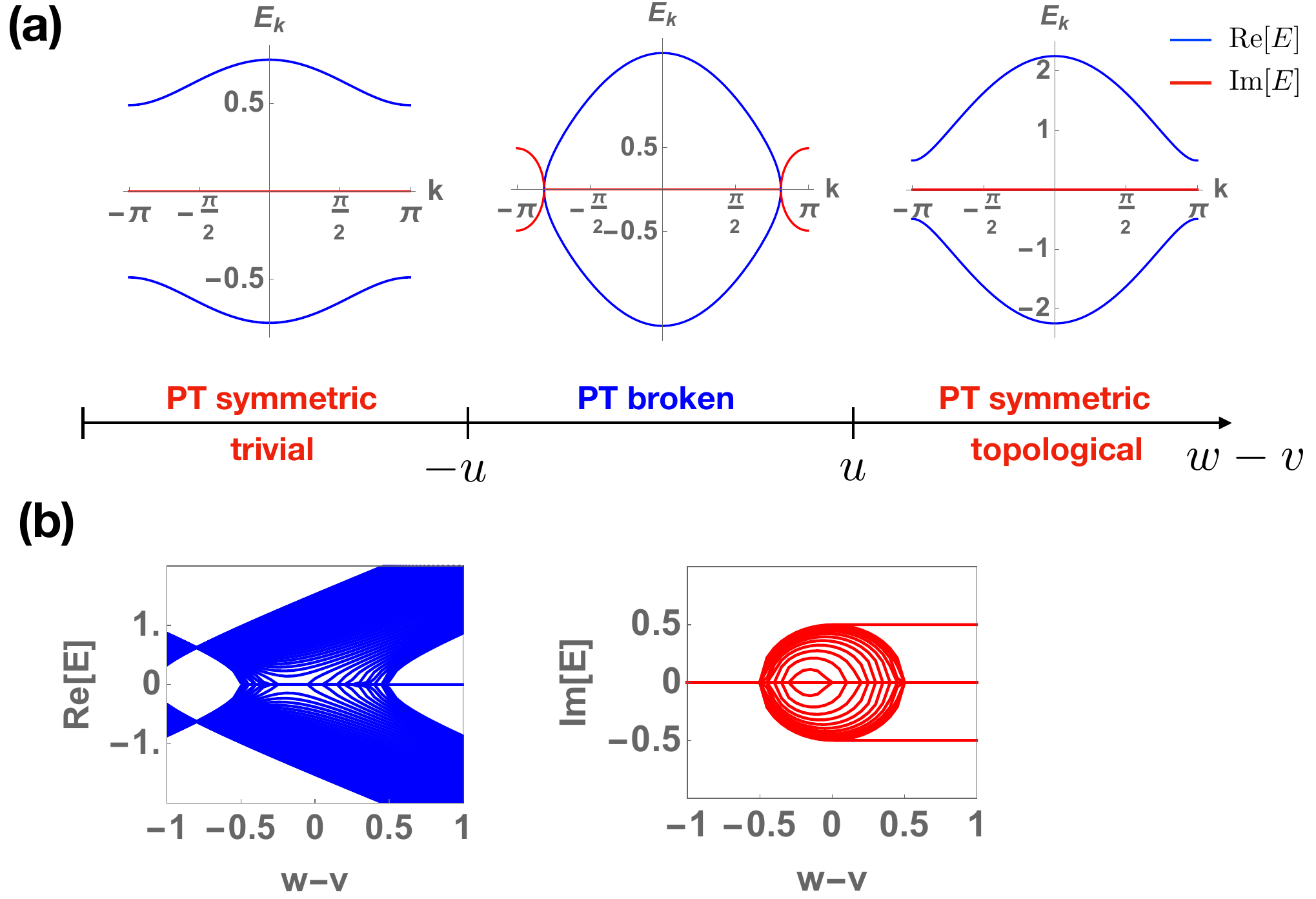}
  \caption{(a) 
    The phase diagram of the non-Hermitian SSH model \eqref{Eq:SSH}
    as a function of $w-v$ with fixed $(u, v)=(0.5, 0.8)$. 
    The upper panels are the corresponding real and imaginary parts of the bulk dispersion.
    (b) The real and imaginary parts of the energy dispersion in the presence of open boundary condition.
    When $w-v>u$, there is a pair of edge modes with imaginary energy $E_{\rm edge}= \pm i u$.}
  \label{F2}
\end{figure}

There are three phases in the non-Hermitian SSH model with $u\neq 0$
[Fig.\ \ref{F2}(a) with the parameters $w,v,u>0$].
In the spontaneously PT-broken phase ($|w-v| <u$),
the energy spectrum is complex and gapless with two exceptional points.
The region with $w-v>u$ realizes one of the PT-symmetric phases,
where the spectrum is fully gapped,
and there is a pair of edge modes with pure imaginary energy 
$E_{\rm edge} = \pm i u$ 
[Fig.\ \ref{F2}(b) (right panel)].
This phase is a topological non-Hermitian phase supporting protected boundary modes 
and characterized by the non-zero global complex Berry phase 
\cite{Liang2013}.
Finally, the other gapped PT-symmetric phase ($w-v<-u$) is trivial
and does not support edge mode [Fig. \ref{F2}(b)]. 

{\it Critical points and (non-unitary) CFT---}
Now we can compute the entanglement entropy and entanglement spectrum
from the correlation matrix or the overlap matrix.  
In this non-Hermitian SSH model with PT-symmetry,
there are two critical points that separate the PT-symmetric phases
and the spontaneously PT-broken phase at $w-v=\pm u$.
When $u=0$, the SSH model has only one critical point at $w=v$
and the system is a critical free-fermion chain. 
There, the entanglement entropy scales logarithmically
in the subsystem size $L_A$,
$S_A \sim ({c}/{3}) \ln L_A$
with $c=1$
\cite{
Calabrese_2004}.

\begin{figure}[t]
  \centering
  \includegraphics[width= 0.5 \textwidth] {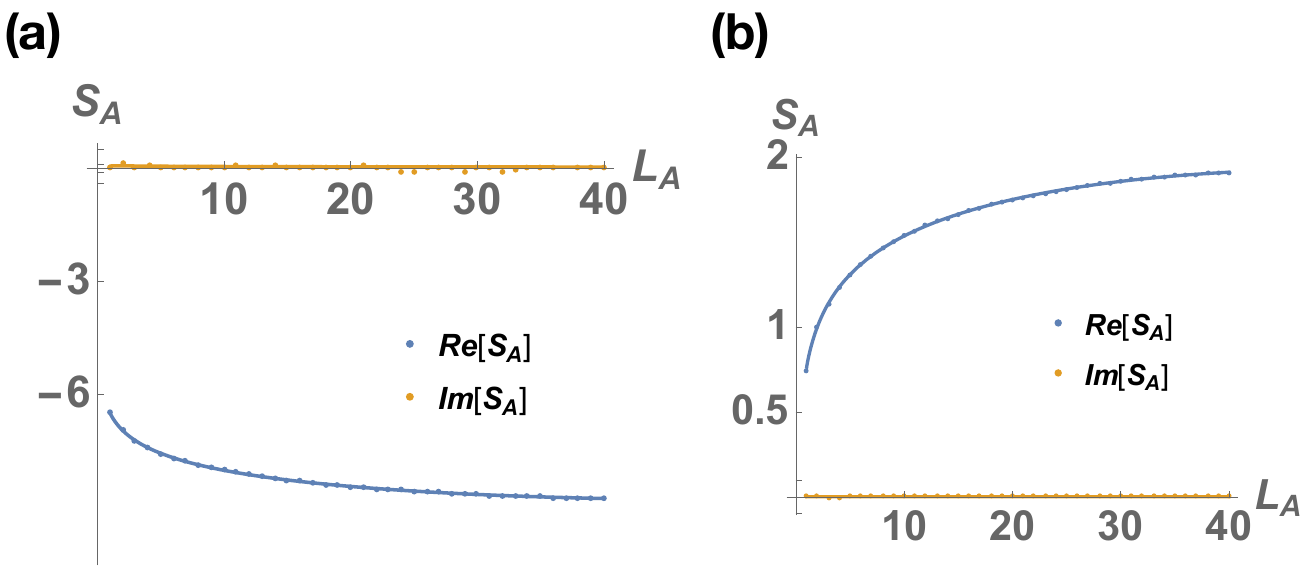}
  \caption{
    The entanglement entropy as a function of the subsystem size $L_A$ 
    with the total system length $L=100$.
    (a) At the critical point separating the trivial PT symmetric gapped phase
    and the PT broken phase,
    $(w,v,u)=(1.3,1.8,0.5)$. 
    The numerical data is fit to
    $S_A (L_A)= -8.81185 -0.666 \ln[\sin({\pi L_A}/{L})]$
    (solid line), leading to the identification $c= -2$.
    (b) In the PT broken phase, $(w, v, u) =(0.8,0.7,0.5)$.
    Here, the state is obtained by filling only negative real-energy modes. 
    The numerical data can be fit to
    $S_A= 1.93+ 0.3337 \ln[\sin({\pi L_A}/{L})]$,
    i.e.,  $c=1$.
  }
  \label{F4}
\end{figure}

For finite $u$, we first analyze the entanglement entropy at the critical point $w-v=-u$
which separates the trivial PT-symmetric phase and the spontaneously PT-broken phase.
At this critical point, we observe that all eigenvalues of the correlation matrix 
are real and they come in pairs,
$\xi_\alpha >1$  and $\xi_\beta = 1-\xi_\alpha<0$.
This pair-wise structure of the eigenvalues guarantees that the entanglement entropy 
is real and negative.
As shown in Fig.\ \ref{F4}(a), we found from numerical calculations that 
the entanglement entropy scales logarithmically 
$S_A = ({c}/{3}) \ln [\sin ({\pi L_A}/{L})] +\rm {const.}$
with the central charge $c=-2$.
Since the spectrum at this critical point is linear around $k=\pm \pi$,
one can write down the effective field theory action as
$S= \int dxdt (
\psi^\dagger_b \bar{\partial} \psi_c+  \bar{\psi}^\dagger_b \partial \bar{\psi}_c)$, 
where 
$\partial = (1/2)(\partial_x - i \partial_t)$,
$\bar{\partial} = (1/2)(\partial_x + i \partial_t)$,
and
$\psi_{b/c}$ ($\bar{\psi}_{b/c})$
represent the fermionic 
fields for the right-moving (left-moving) modes.
(We have set the Fermi velocity to be one).
We identify $\psi^\dagger_{b(c)}$ as the right(left) creation operator  $\psi^\dagger_{R(L)}$ defined in 
Eq.\ (\ref{Eq1}).
A crucial observation here is that states associated with 
these right and left fermionic operators are not normalizable at the critical point
\cite{Norm}:
The ill-defined norm of the quantum states can be thought of as a hallmark of the ghost theory.
The above action, with proper assignment of the conformal dimensions,
defines the {\it bc}-ghost CFT with central charge $c=-2$ 
\cite{SM, Polchinski,Blumenhagen,FRIEDAN198693, KAUSCH2000513, Kausch1995,GURUSWAMY1998661}.
The entanglement entropy detects
the correct central charge as we expect.

The appearance of the negative central charge, however, 
is sensitive to the choice of the boundary condition.
To obtain the $c=-2$ scaling of the entanglement entropy,
we need to choose the periodic boundary condition and consider the half-filled ground state,
where we fill the (left/right) state at the crossing point but with a tiny momentum shift \cite{numrics}.
On the other hand, imposing anti-periodic, open boundary conditions, 
or simply removing the state at the crossing points $k=\pm \pi$, 
we recover the central charge $c=1$ in the entanglement entropy scaling
\cite{HS}.
The sensitivity to boundary conditions is consistent with the fact that 
the ill-defined norm occurs only at the crossing point $k=\pm \pi$.
Furthermore, we monitor the entanglement entropy scaling as a function of
the twisting angle of the boundary condition (i.e., a shift of single-particle momenta) for
very small ($\sim 10^{-7}$) to large twisting angle ($\sim 2\pi$).
We observe that the entanglement entropy scaling crosses over 
from a convex ($c=-2$) to concave function ($c=1$)
\cite{SM}.
This crossover also occurs by going from the critical point to a PT-symmetric phase with a small gap.  
We note that a similar crossover was also observed
in the quantum Ising chain in an imaginary magnetic field
\cite{Dupic2018}.

We now turn to the other critical point ($w-v=u$) separating 
the topological PT-symmetric phase and the spontaneously PT-broken phase.
It exhibits a very different entanglement entropy scaling. 
At this critical point, we observe an additional pair of eigenvalues 
$\xi_{\pm, \alpha} = 0.5 \pm i I_\alpha$,
where $ I_\alpha$ depends on the subsystem size.
This pair in the entanglement spectrum mimics
the protected boundary modes in the gapless point in the physical spectrum 
[see Fig.\ \ref{F2}(b) at $w-v=u$ and Fig.\ \ref{ES_EE}(a) in \cite{SM}].
Although this pair of eigenvalues 
contributes to the pure real part of the entanglement entropy,
the imaginary part 
$I_\alpha$ depends on the total system size.  
We find the entanglement entropy
shows the scaling
$S_A = \alpha \ln [\sin 
({\pi L_A}/{L})]+\rm {const.}$,
where $\alpha$ decreases as a function of the total system size 
[Figs.\ \ref{ES_EE}(b-c) in \cite{SM}]. 
The length-dependent coefficient $\alpha$ is not described by CFT.
However, these two critical points are seemingly described by
the same effective field theory, and one would expect the same scaling behavior in the entanglement entropy.

To shed some light on this issue, we consider another bipartition which cuts through the unit cell\cite{2006PhRvB..73x5115R}.
With this shifted unit-cell, the hopping $w$ and $v$ are interchanged and the
entanglement spectra are identical to the critical point that separates the
trivial PT-symmetric phase and the spontaneously PT-broken phase.
We recover the $c=-2$ entanglement entropy scaling under this bipartition.
The above story reminds us of the protected boundary modes
(both in the physical and entanglement spectra)
in the symmetry-enriched critical point in the Hermitian system discussed in Ref.\ \cite{Verresen2019}.
There, by 
redefining the unit-cell, we also recover the $c=1$ scaling of
the free-fermion system \cite{SM}.
In this sense, the non-unitary CFT appearing at the critical point separating
the topological PT-symmetric phase and the spontaneously
PT-broken phase may be viewed as a non-Hermitian analog of symmetry-enriched critical theory.

Lastly, let us consider the entanglement properties in the spontaneously PT-broken phase $(|w-v|<u)$,
where there are two exceptional points.
In this phase, some of the single-particle energies are purely imaginary and filling such single-particle states would be unphysical.
Here, we consider the "ground state" which is constructed by filling only states with real energy.
As shown in Fig.\ \ref{F4}(b), the entanglement entropy of this "ground state" follows the CFT scaling behavior, 
$S_A = ({c}/{3}) \ln [\sin ({\pi L_A}/{L})]+\rm {const.}$
with $c=1$.  
The appearance of the $c=1$ CFT behavior can be understood as follows.
At the exceptional points, two eigenstates are coalescing into one and the
Hamiltonian cannot be diagonalized.
One can expand the Hamiltonian at the exception point $k_{\rm EP}$ with a Jordon block form
\cite{Kanki2017,Demange_2011}, 
$H=\int dk\, \Psi^\dagger_k
\left[\sigma_z(k_{\rm EP}-k) +\gamma (\sigma_x + i \sigma_y) \right] \Psi_k$
where $\Psi^\dagger_k= (\psi^\dagger_{1,k},\psi^\dagger_{2,k})$ are fermionic fields,
and $\gamma$ is an arbitrary complex number.
(Once again, we set the Fermi velocity to be unity \cite{unit}.)
At the exceptional point $k_{\rm EP}$, two eigenstates collapse to one eigenstate $(1,0)$ with energy $-(k-k_{\rm EP})$.
The ground state can be expressed as $|G\rangle = \prod_{k-k_{\rm EP}>0} (1,0)^{\rm T}$, which has the identical form of the ground state
of the free Dirac theory with 
central charge $c=1$.

\begin{figure}[t]
  \centering
  \includegraphics[width=0.5\textwidth] {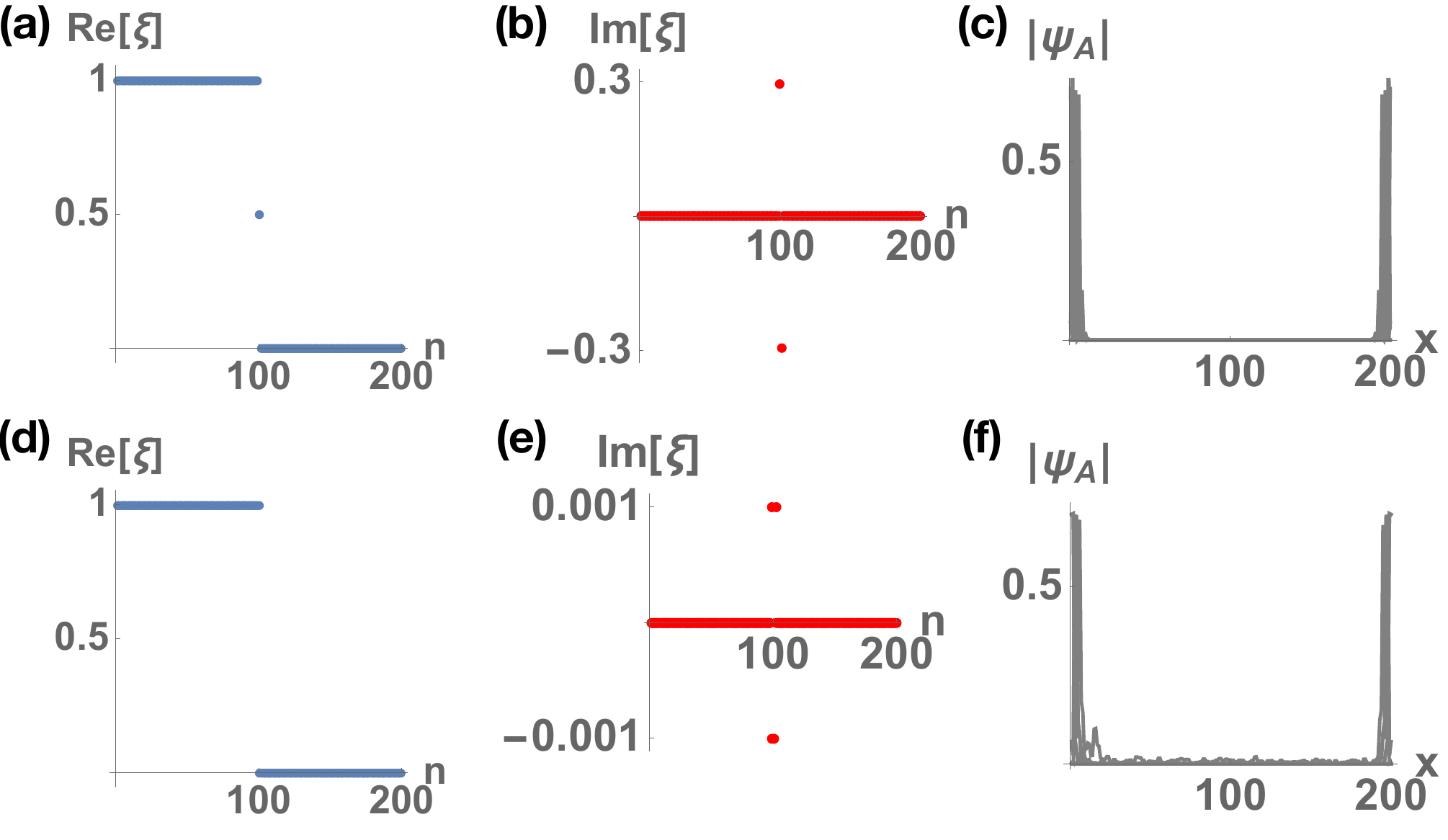}
  \caption{
    The real part (a) and imaginary part (b) of the eigenvalues of the
    correlation matrix $C_A$ with $(w-v, u)=(0.9, 0.5)$.
    There are two mid-gap states.
    (c) The wavefunction amplitude of the localized states including two mid-gap states.
    The real part (d) and imaginary part (e) of the eigenvalues of the
    correlation matrix $C_A$ with $(w-v, u)=(-0.9, 0.5)$.
    There are four states with imaginary eigenvalues.
    (f) The wavefunction amplitude of the localized states where the total system is two hundreds sites and $n$ is the eigen-level index.}
  \label{F3}
\end{figure}

{\it Fully gapped phases and entanglement spectrum---}
Next, we study the PT-symmetric, fully gapped phases, where the ground state is well defined since there
is no imaginary energy mode.
Once again, the entanglement spectrum is obtained from the eigenvalues of the correlation matrix and the overlap matrix
with the periodic boundary condition. 
In the topological non-Hermitian phase ($w-v>u$) where the physical edge modes are present, 
there are two mid-gap states in the entanglement spectrum with
${\rm Re}\, [\xi] =1/2$
and non-vanishing imaginary part ${\rm Im}\, [\xi] \neq 0$ 
[Fig.\ \ref{F3} (a-b)].
In addition to these two mid-gap states localized at the entangling boundaries,
there are numerous other localized boundary modes which are not the mid-gap states in the correlation matrix  
[Fig.\ \ref{F3} (c)].
On the other hand, in the trivial phase where no physical edge modes 
are present,
we observe four additional states with non-vanishing imaginary eigenvalues of the correlation matrix 
[Fig.\ \ref{F3} (d-e)].
There are also numerous localized boundary modes which are not the mid-gap states in this phase.
These localized modes in the correlation matrix are similar to the non-Hermitian skin effect in non-Hermitian 
systems with the open boundary condition
\cite{Xiong2018,Yao2018,Alvarez2018}.

\subsection{Non-Hermitian Chern insulators}

Finally, we consider the non-Hermitian Chern insulator~\cite{Kawabata2018} in two dimensions 
defined in momentum space by
\begin{align}\label{CI_Hamiltonian1}
\mathcal{H}(\mathbf k)=&(m + t \cos k_x + t \cos k_y )\sigma_x\nonumber\\
&+(i \gamma + t \sin k_x )\sigma_y + (t \sin k_y )\sigma_z,
\end{align}
where $t, m,$ and $\gamma$ are real parameters and we assume $t > 0$.
The complex dispersion relation is given by
$
E_{\pm}(\mathbf k) = \pm[(m + t \cos k_x + t \cos k_y )^2
+\sqrt{(i \gamma + t \sin k_x )^2 + (t \sin k_y )^2}.
$
Exceptional points appear when the two bands satisfy $E_{\pm}(\mathbf k_{\textmd{EP}})=0$ at certain momenta
$\mathbf k=\mathbf k_{\textmd{EP}}$, at which 
the Hamiltonian~\eqref{CI_Hamiltonian1} is defective and the eigenstates
coalesce and linearly depend on each other.
The gapped phases with $E_{+}(\mathbf k) \neq E_{-}(\mathbf k)$ for all $\mathbf k$ are
characterized by the first Chern number,
and in the case of $\gamma = 0$, the model reduces to a Hermitian Chern insulator.
We computed the entanglement spectrum of the non-Hermitian Chern insulator model
\eqref{CI_Hamiltonian1}
from the correlation matrix and/or the overlap matrix with the periodic boundary conditions in both $x$ and $y$-directions.
Since it has been known that the existence of the bulk-edge correspondence in (physical) non-Hermitian systems
is sensitive to boundary conditions,
in the following we investigate the entanglement spectrum of the non-Hermitian gapped phases by setting entangling boundary either in the
$y$ direction or in the $x$ direction. 

\begin{figure}[t]
  \centering
  \includegraphics[width= 0.5 \textwidth] {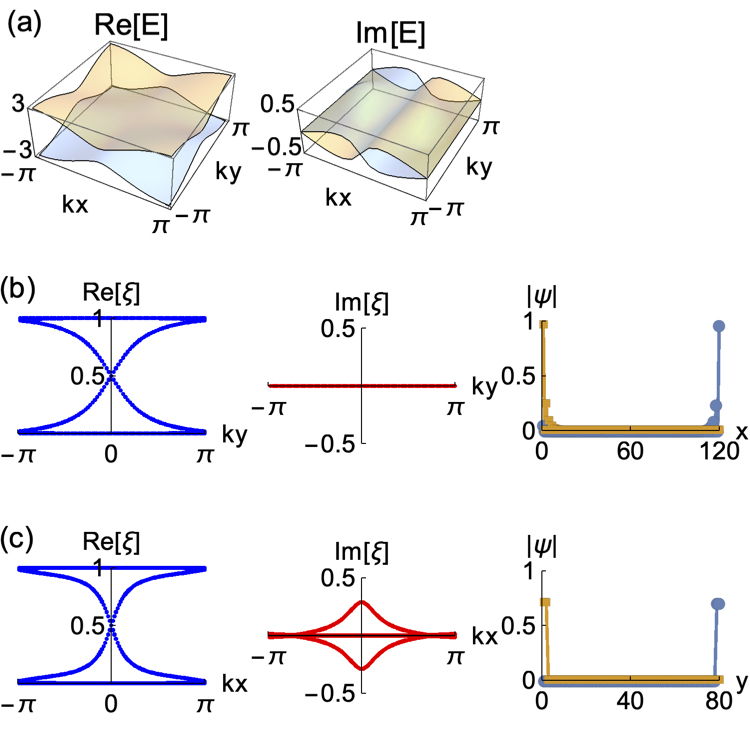}
  \caption{
    Topologically nontrivial gapped phases with nonzero Chern number
    $C = -1$ ($t = 1.0, m = -1.0, \gamma = 0.5$). (a) Complex-band structures of the non-Hermitian Chern
    insulator $E_{\pm} = E_{\pm}(k_x, k_y)$, where the orange and blue bands
    represent $E_{+}$ and $E_{-}$, respectively.
    (b) Complex entanglement spectrum as a function of $k_y$,
    and the mid-gap modes at $k_y=0$ and $\xi=0.5$ along $x$ direction.
    (c) Complex entanglement spectrum as a function of $k_x$,
    and the mid-gap modes at $k_x=0$ for $\xi=0.5 - 0.289 i$ ~(yellow squares) and for $\xi=0.5 + 0.289 i$~(gray dots) along $y$ direction.}
  \label{Fig_Chern}
\end{figure}

In the topologically non-trivial gapped phase,
we observe two chiral mid-gap modes localized at the right and left edges, respectively 
[Figs.~\ref{Fig_Chern} (b) and (c)]. 
In particular, for the entangling boundary running along the $x$-direction,
the right mid-gap mode has the largest positive imaginary part for $\gamma> 0$,
whereas the left mid-gap mode has the largest negative imaginary part
[Fig.~\ref{Fig_Chern} (c)].
In contrast, the imaginary parts of the mid-gap states vanish for the entangling boundary
along the $y$-direction
[Fig.~\ref{Fig_Chern} (b)].
This phenomenon implies the amplification of the right mid-gap mode and the attenuation of the left mid-gap mode, 
similar to the behaviors of the physical edge modes in the non-Hermitian Chern
insulator and a topological insulator laser discussed in other contexts~\cite{Xu2017,Harari2018,Kawabata2018}.

\section{Conclusion and outlook}

Many critical points/phases in Hermitian quantum many-body systems are described by unitary CFTs.
For example, the Tomonaga-Luttinger liquid phase with an integral central charge is rather ubiquitous
in one dimensional many-body systems.
In contrast, virtually nothing has been known on the many-body aspects of critical points/phases
appearing in non-Hermitian systems, in particular in terms of their (conformal) field theory description.
In this letter, from the entanglement entropy scaling, we identified the non-unitary CFT with $c=-2$
in a PT-symmetric non-Hermitian hosting topological phases. 
It deserves further investigation as to why this particular CFT is realized.
We also demonstrated the entanglement spectrum can detect the topological properties 
in the gapped phase of the PT-symmetry SSH mode and the non-Hermitian Chern insulator.

{\it Note added}--
During the preparation of this manuscript, we became aware of
a partially related work by Herviou et al. \cite{Herviou2019_2}.



\section{Acknowledgments}
P.-Y.C. thanks Natan Andrei, Pochung Chen, Markus Heyl, Yi-Ping Huang, Elio K\"{o}nig, Yashar Komijani, Sarang Gopalakrishnan, and Inti Sodemann for valuable discussions.
We are thankful to Benoit Estienne for discussion about extracting the real and effective central charges, and thankful to
Hassan Shapourian for pointing out the scaling of the entanglement entropy is sensitive for the boundary conditions. P.-Y.C. was supported by the Young Scholar Fellowship Program by Ministry of Science and Technology (MOST) in Taiwan, under MOST Grant for the Einstein Program MOST 108-2636-M-007-004.
X.W. is supported by the Gordon and Betty Moore Foundation’s EPiQS initiative through Grant No. GBMF 4303 at MIT. 
S.R. is supported in part by the National Science Foundation under
Grant No. DMR-1455296, 
and by a Simons
Investigator Grant from the Simons Foundation.

\bibliographystyle{apsrev4-1}
\bibliography{references}

\begin{thebibliography}{67}
\bibitem [{Nor()}]{Norm}%
  \BibitemOpen
  \href@noop {} {\bibinfo  {journal} {The norm of a state $|c\rangle $ with
  $\psi_c |c\rangle =0 $ is the left eigenvector defined in Eq. (\ref{Eq:L}),
  $|c\rangle = (-v_k/|v_k| \sin^* (\phi_k/2), \cos^* (\phi_k/2))^{\rm T}$ with
  $\phi_k = \tan^{-1}[ -i |v_k|/u]$ being purely imaginary. The norm of this
  state $\langle c| c \rangle= \sin (\phi_k/2 ) \sin (\phi_k^*/2 ) + \cos
  (\phi_k/2 ) \cos (\phi_k^*/2 )$ diverges at the critical point $v-w=u$, which
  is ill-define}\ }\BibitemShut {NoStop}%
\bibitem [{num()}]{numrics}%
  \BibitemOpen
\bibfield  {journal} {  }\href@noop {} {\bibinfo  {journal} {The (left/right)
  state is divergent at the crossing point ($k=\pi$). When computing the
  entanglement entropy, we shift the momentum by a small small amount $\delta =
  0.0000001$ which avoids the divergence at the crossing point.}\ }\BibitemShut
  {NoStop}%
\bibitem [{uni()}]{unit}%
  \BibitemOpen
  \href@noop {} {\bibinfo  {journal} {The velocity at the exceptional point is
  infinity. However, the central charge will not depend on the velocity of the
  effective theory, we take the velocity to be unity}\ }\BibitemShut {NoStop}%
\end{thebibliography}













\clearpage

\pagebreak

\newpage

\begin{widetext}
\begin{center}
\noindent{\Large Supplementary Material for ``Entanglement spectrum and entropy in topological non-Hermitian systems and non-unitary conformal field theory".}
\end{center}


\section{Correlation Matrix}

In this supplementary material,
we present 
the detailed derivations of 
the entanglement properties from the correlation matrix and the overlap matrix.

\label{App:CM}
We first use the correlation matrix
\cite{Peschel2009} to extract eigenvalues of the reduced density matrix in a free fermion system.
Since the theory is free, the reduced density matrix has a Gaussian form $\rho_A = \frac{1}{\mathcal{N}}\exp (- \sum_{\alpha,\beta}\mathcal{H}_{\alpha\beta}^E \phi^\dagger_{\alpha} \phi_\beta)$,
where $\mathcal{H}_{\alpha\beta}^E$ refers to the entanglement Hamiltonian.
The correlation matrix is defined as
\begin{align}
C_{ij} = \langle G_L |  \phi_i^\dagger \phi_j  |G_R\rangle ={\rm Tr}  \rho_A  \phi_i^\dagger \phi_j,
\end{align} 
where $|G_R\rangle$ and $|G_L\rangle$ are the right and left ground states. 
One can simultaneously diagonalize $C_{ij}$ and $\mathcal{H}^E_{\alpha\beta}$
and 
find $C=\frac{e^{-\mathcal{H}^E}}{1+e^{-\mathcal{H}^E}}$.
The correlation matrix can be expressed by the left and right eigenvectors as
\begin{align}
C_{ij} &= \langle G_L |  \phi_i^\dagger \phi_j  |G_R\rangle \notag \\
&= \langle 0 |    \prod_{i' \in {\rm occ.}} (\sum_\alpha L_{\alpha i'} \phi_\alpha ) \phi_i^\dagger     \phi_j \prod_{j'} (\sum_\beta R^\dagger_{\beta j'} \phi_\beta^\dagger)                            |0\rangle.
\end{align} 
By using the commutation relation $\{\phi_i, \phi_j^\dagger \} =\delta_{ij}$ and $\phi_i|0 \rangle=0$,
we have
\begin{align}
 \phi_j \prod_{j'} (\sum_\beta R^\dagger_{\beta j'} \phi_\beta^\dagger)  |0\rangle &= R^\dagger_{1 j} \prod_{j' \neq 1} \psi_{Rj'}^\dagger |0 \rangle - R^\dagger_{2 j} \prod_{j' \neq 2} \psi_{Rj'}^\dagger |0 \rangle
 + R^\dagger_{3 j} \prod_{j' \neq 3} \psi_{Rj'}^\dagger |0 \rangle-\cdots \notag\\
  \phi_i \prod_{i'} (\sum_\alpha L^\dagger_{\alpha i'} \phi_\alpha^\dagger)  |0\rangle &= L^\dagger_{1 i} \prod_{i' \neq 1} \psi_{Li'}^\dagger |0 \rangle - L^\dagger_{2 i} \prod_{i' \neq 2} \psi_{Li'}^\dagger |0 \rangle
 + L^\dagger_{3 i} \prod_{i' \neq 3} \psi_{Li'}^\dagger |0 \rangle-\cdots, 
 \label{Eq:rel}
\end{align}
where $ \psi_{Rj'}^\dagger = \sum_{\beta} R^\dagger_{\beta j'} \phi_\beta^\dagger$ and $ \psi_{Li'}^\dagger = \sum_{\alpha} L^\dagger_{\alpha i'} \phi_\alpha^\dagger$,
which satisfy $\{ \psi_{Li}, \psi_{Rj}^\dagger \} =\delta_{ij}$.
Using Eq.\ (\ref{Eq:rel}), the correlation matrix is
\begin{align}
C_{ij} =\sum_{\alpha} L_{\alpha i} R^{\dagger}_{\alpha j}.
\end{align}
One should notice that $C^*_{ji} = \sum_{\alpha} R_{\alpha i} L^{\dagger}_{\alpha j} \neq C_{ij}$ is not Hermitian.
The entanglement entropy is
\begin{align}
S_A &= - {\rm Tr} \rho_A \ln \rho_A =  - {\rm Tr} [ \frac{e^{-\mathcal{H}^E}}{1+e^{-\mathcal{H}^E}} \ln \frac{e^{-\mathcal{H}^E}}{1+e^{-\mathcal{H}^E}} +
 \frac{1}{1+e^{-\mathcal{H}^E}} \ln \frac{1}{1+e^{-\mathcal{H}^E}} ]  \notag\\
 &= -\sum_{\delta} \left[ \xi_\delta \ln \xi_\delta +(1-\xi_\delta) \ln (1-\xi_\delta)  \right],
\end{align}
where $\xi_\delta$ are the eigenvalues of $C_{ij}$.

\section{Overlap Matrix}
\label{App:OM}
Alternatively, we can use the overlap matrix to extract the entanglement
properties.
The basic idea is to rotate the reduced density matrix in a 
new biothogonal basis in the subsystem $A$,
$\{|L^{A(B)}_i\rangle , |R^{A(B)}_i\rangle\}$ such that $\langle L^{A(B)}_i | R^{A(B)}_i \rangle = \delta_{ij}$.
The rotation matrix is constructed from the overlap matrix
\cite{Klich2006},
\begin{align}
M^A_{\alpha\beta} = \int_{x\in A} dx \langle L_\alpha| R_\beta\rangle  = \sum_i p_i   (L^A_{i \beta} )^\dagger R^A_{i\alpha},  
\end{align}
where $L^A_{i \beta}$ and $R^A_{i\alpha}$ are the corresponding left and right eigenvectors of $M^A$ with eigenvalues $p_i^*$ and $p_i$.
Notice that the overlap matrix is non-Hermitian,
\begin{align}
[M^A_{\alpha\beta}]^\dagger= [M^A_{\beta\alpha}]^* = [ \int_{x\in A} dx \langle L_\beta | R_\alpha\rangle]^*
\neq  \int_{x\in A} dx \langle L_\alpha  | R_\beta \rangle.
\end{align}
So in the diagonal basis, the non-Hermitian  overlap matrix is
\begin{align}
\sum_{\alpha \beta}(L^A_{i\alpha})^\dagger M^A_{\alpha \beta} R^A_{\beta i }= p_i.
\end{align}
Thus we can normalized the left and right eigenvectors of $M^A$ as
\begin{align}
\sum_{\alpha \beta} \frac{(L^A_{i\alpha})^\dagger}{\sqrt{p_i}}  \langle \tilde{L}^A_{\alpha}|\tilde{R}^A_{\beta}\rangle \frac{ R^A_{\beta j }}{\sqrt{p_j}}= \delta_{ij}.
\label{Eq: LBO}
\end{align}
Here we express $M^A_{\alpha \beta} =  \langle \tilde{L}^A_{\alpha}|\tilde{R}^A_{\beta}\rangle$,
where $| \tilde{L}^A_{\alpha}\rangle$ and $| \tilde{R}^A_{\alpha}\rangle$ are left and right vectors span only in subsystem $A$.
These left and right vectors  are not yet biorthogonal,
but from Eq.\ (\ref{Eq: LBO}), we can construct the local biorthogonal basis as
\begin{align}
|R^A_i\rangle &= \sum_\alpha  \frac{R^A_{i\alpha}}{\sqrt{p_i}}  | \tilde{R}^A_{\alpha} \rangle,  \quad 
|L^A_i\rangle = \sum_\alpha  \frac{L^A_{i\alpha}}{\sqrt{p_i}}  | \tilde{L}^A_{\alpha} \rangle, 
\quad 
\langle L^A_i |R^A_j\rangle = \delta_{ij}.
\end{align}
From the property of the overlap matrix $M^A+M^B =\mathbb{I}$, $M^A$ and $M^B$ can be simultaneously diagonolized 
with the corresponding eigenvalues $p_i$ and $1-p_i$,
\begin{align}
\sum_{\alpha \beta}(L^A_{i\alpha})^\dagger M^A_{\alpha \beta} R^A_{\beta j } + \sum_{\alpha \beta}(L^A_{i\alpha})^\dagger M^B_{\alpha \beta} R^A_{\beta j } 
= p_i \delta_{ij} + (1- p_i)  \delta_{ij}= \delta_{ij}.
\end{align}
Thus the biorthogonal basis $\{|R^B_i\rangle, |L^B_i\rangle \}$ in subsystem $B$ can also be constructed from the rotation matrices $R^A_{\alpha i } $ and $L^A_{\beta j }$. 
The right vector $|R_\alpha\rangle$in the total system can be rotated by $R^A_{\alpha i } $ such that 
\begin{align}
\sum_{\alpha} R^A_{\alpha i }  |R_\alpha\rangle = \sqrt{p_i} |R^A_i\rangle + \sqrt{1- p_i} |R^B_i\rangle.
\end{align}
Now we consider many-body wave function after the rotation
\begin{align}
|\tilde{G}_R\rangle = \prod_{i} (\sqrt{p_i} {\psi^A}^\dagger_{Ri} +\sqrt{1-p_i} {\psi^B}^\dagger_{Ri} )   |0\rangle,
\end{align}
where $ {\psi^{A(B)}}^\dagger_{Ri}  |0\rangle = |R^{A(B)}_i \rangle$.
Then the density matrix in the rotated many-body wave function has a tensor product form
\begin{align}
\rho= |\tilde{G}_R\rangle \langle \tilde{G}_L | = \bigotimes_i | \nu_{Ri} \rangle \langle \nu_{Li} |, 
\end{align}
where $| \nu_{Ri}\rangle = \sqrt{p_i} |R_i^A\rangle_A | 0\rangle_B + \sqrt{1-p_i} |0\rangle_A | R_i^B\rangle_B$.
Then the reduced density matrix has a tensor product form
\begin{align}
\rho_A 
= \sum_i \langle L^B_{i} |\rho | R^B_{i} \rangle
=  \bigotimes_i (p_i |L^A_i   \rangle \langle R^A_i | +(1-p_i) |0\rangle \langle 0 |).
\end{align}

Now we can compute the entanglement entropy as
\begin{align}
S_A = - \sum_i [p_i \ln p_i + (1-p_i) \ln (1-p_i)].
\end{align}
Furthermore, if we define $p_i:=\frac{q_i}{{q_i+q_i^{-1}} }$ and $1-p_i:=\frac{q^{-1}_i}{{q_i+q_i^{-1}} }$,
the entanglement entropy is
\begin{align}
S_A &= - \sum_i [\frac{q_i}{{q_i+q_i^{-1}} } \ln \frac{q_i}{{q_i+q_i^{-1}} } + \frac{q^{-1}_i}{{q_i+q_i^{-1}} } \ln \frac{q^{-1}_i}{{q_i+q_i^{-1}} }] \notag\\
& =  \sum_i [\ln (q_i + q_i^{-1})-\frac{q_i-q_i^{-1}}{q_i + q_i^{-1}} \ln q_i].
\end{align}
The eigen-spectrum of $\rho_A$ gives
the entanglement spectrum.

\section{Crossover of the entanglement entropy scaling from $c=-2$ to $c=1$}

\begin{figure}[htbp]
\centering
\includegraphics[width= 0.7 \textwidth] {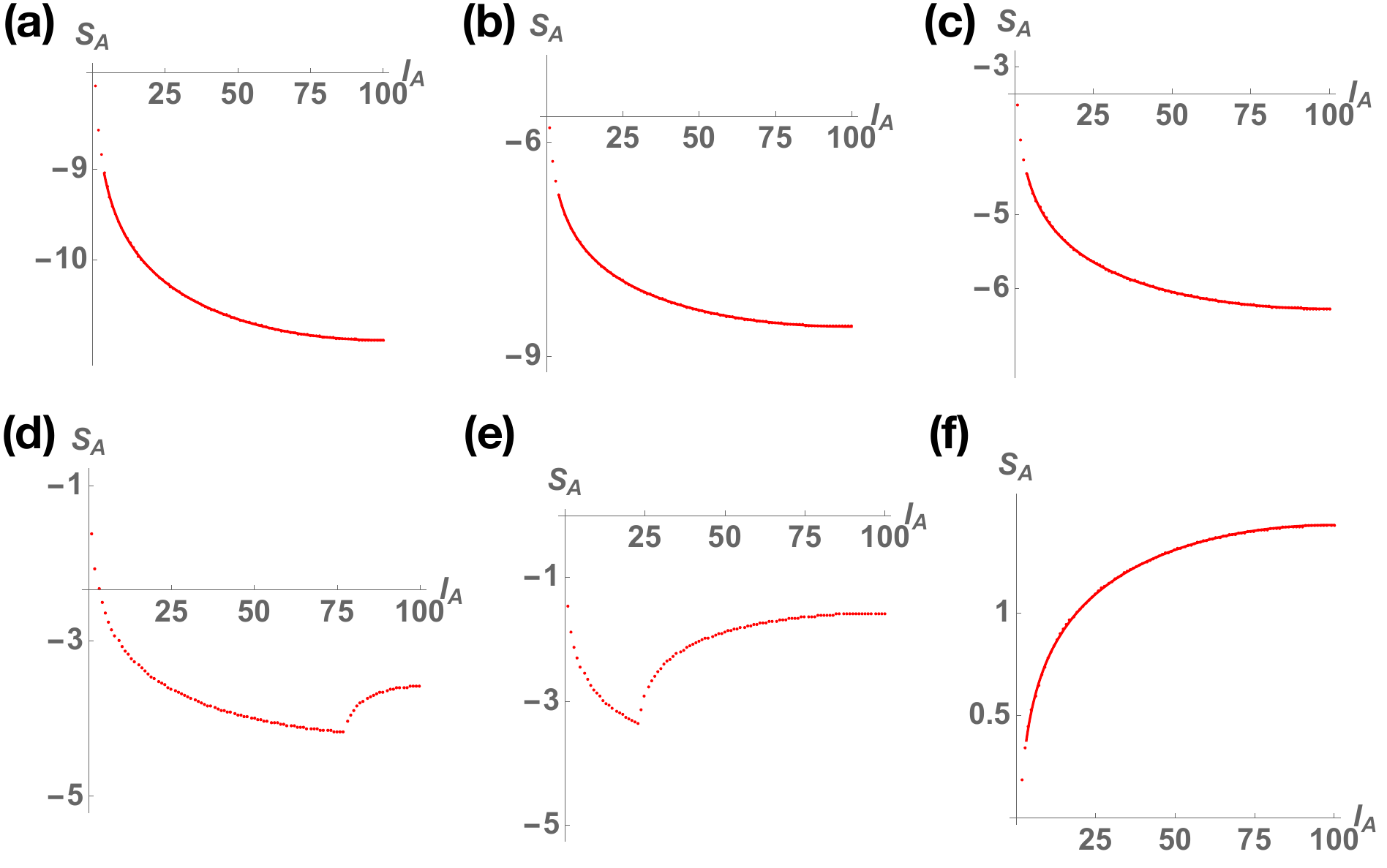}
\caption{The entanglement entropy scaling as a function of subsystem size $l_A$ with fixed total system size $L=200$ and different momentum shifts $\delta$. (a) $\delta=0.000001$, (b) $\delta=0.00001$, (c) $\delta=0.0001$, (d) $\delta=0.0008$, (e) $\delta=0.001$, (f) $\delta=0.01$. The corresponding $c/3$ for (a),(b),(c), and (f) are $0.666$,$0.666$,$0.660$ and $0.346$, respectively. }
\label{ESS_delta}
\end{figure}

\begin{figure}[htbp]
\centering
\includegraphics[width= 0.5 \textwidth] {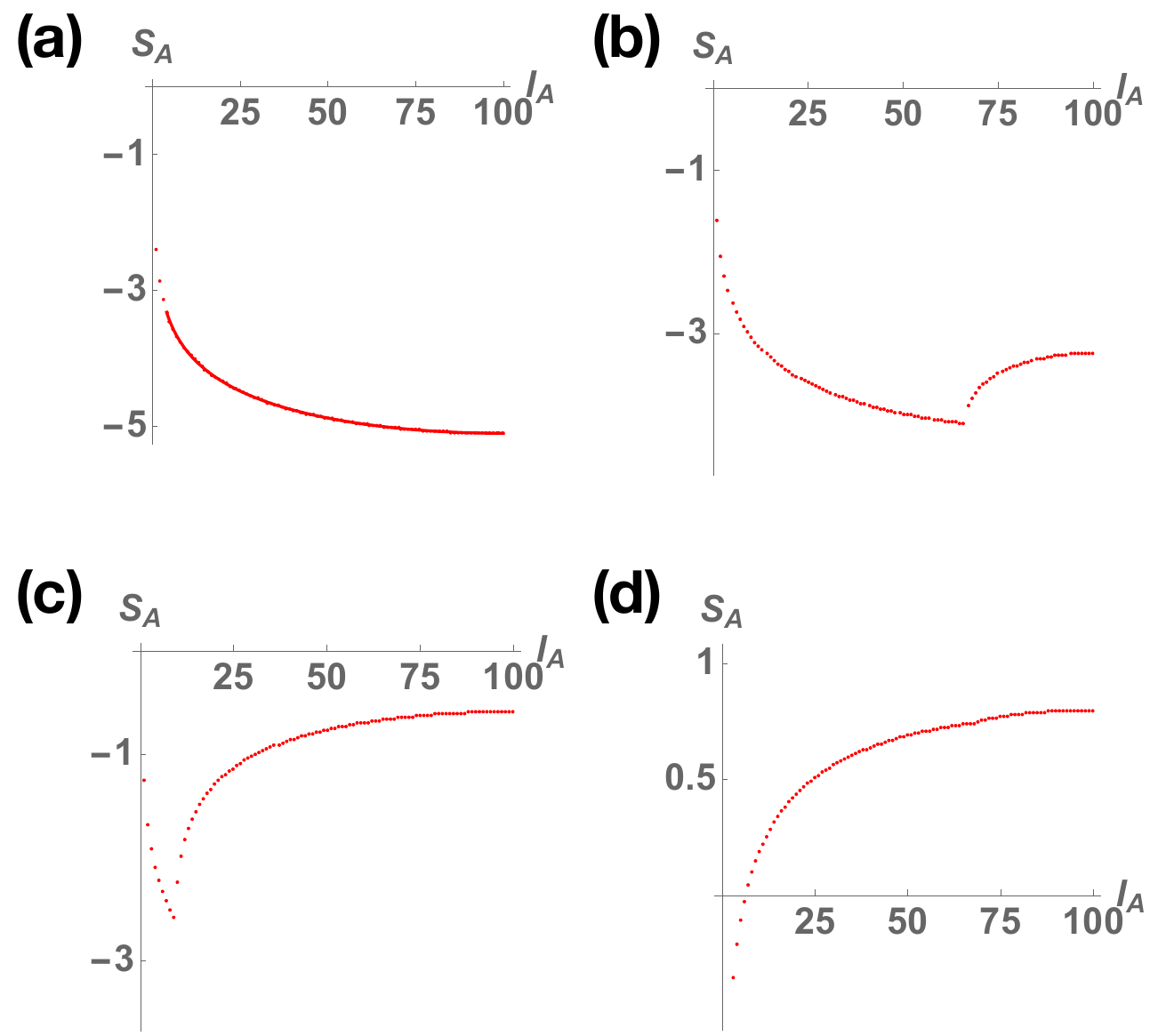}
\caption{The entanglement entropy scaling as a function of subsystem size $l_A$ with fixed total system size $L=200$, momentum shift $\delta=0.000001$, and different gap $\Delta = 2 \sqrt{|v_k|^2 - u^2}$. Here we fix $(w,v)=(1.3,1.8)$ and change $u$. (a) $\Delta=0.001$, (b) $\Delta=0.0025$, (c) $\Delta=0.004$, and (d) $\Delta=0.01$. }
\label{ESS_gap}
\end{figure}

As we mentioned in the main text, both left and right eigenvectors are singular
at the crossing point ($k=\pi$). To get the $c=-2$ entanglement entropy scaling
behavior, we introduce a tiny momentum shift $\delta$ to avoid
directly taking the crossing point.

We observe that if the momentum shift is comparable to the size of the discrete momentum $2\pi/L$, the entanglement entropy scaling gives the central charge $c=1$.
The crossover from $c=-2$ to $c=1$ entanglement entropy scaling is shown in Fig.
\ref{ESS_delta}.
The entanglement entropy as a function of the subsystem size goes from a convex
function to a concave function.
This crossover behavior can also be seen when
we are slightly away from the critical point [\ref{ESS_gap}].
In other words, a small finite momentum shift can be seen as introducing a small gap at the crossing point.
We can linearize the spectrum at the crossing point and 
the corresponding velocity is $v_{\rm eff.} = \sqrt{w v}$.
We can convert the momentum shift to the corresponding gap
$\Delta \sim 2 v_{\rm eff.}  \delta$ which is the same order of the crossover behavior shown in Fig.\ \ref{ESS_gap}.
This crossover is also observed in the quantum Ising chain in an imaginary
magnetic field
\cite{Dupic2018}.

\section{Entanglement properties
  at the critical point separating
  the topological PT symmetric phase and the
  spontaneously PT broken phase}

In the SSH model with PT symmetry,
the critical point which separates
the topological PT symmetric phase and the PT broken phase
has a pair of "boundary modes" in the entanglement spectrum
$\xi_{\pm, \alpha}=0.5 \pm i  I_\alpha$ [Fig.\ \ref{ES_EE}(a)].
Since $\xi_{\pm, \alpha}$ and $\xi_{\mp, \alpha}=1-\xi_{\pm, \alpha}$
are complex conjugate to each other,
this pair does not generate the imaginary part of the entanglement entropy,
i.e., the entanglement entropy is still real.
The scaling behavior of the entanglement entropy
is $S_A = \alpha(L) \ln [\sin[\frac{\pi l_A}{L}]] + {\rm const.}$
as shown in Fig.\ \ref{ES_EE}(b).
The coefficient $\alpha(L)$ depends on the total system size $L$
as shown in Fig.\ \ref{ES_EE}(c).
At thermodynamic limit $1/L \to 0$, we expect
 this coefficient vanishes.

\begin{figure}[htbp]
\centering
\includegraphics[width= 0.9 \textwidth] {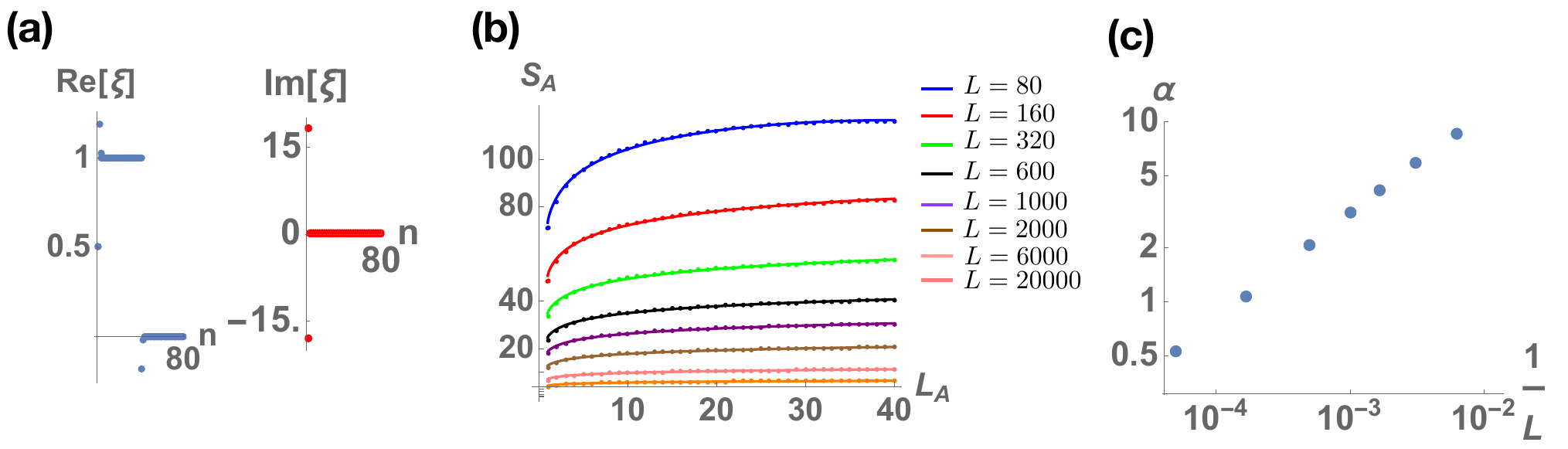}
\caption{(a) Entanglement spectrum at the critical point which separates the topological PT symmetric phase
and the PT broken phase. The parameters are  $(w,v,u)=(1.8,1.3.0,5)$. (b) The entanglement entropy as a function of the subsystem size $L_A$
with different total system size $L$. The scaling of the entanglement entropy satisfies $S_A (L_A)= \alpha(L)  \ln[\sin(\frac{\pi L_A}{L})] + {\rm const.} $ with $\alpha(L)$ depending on the total system size. 
(c) The log-log plot of the $\alpha(L)$ as a function of $1/L$. }
\label{ES_EE}
\end{figure}

The "boundary modes" in the entanglement spectrum are not exponentially
localized but show power-law decay at the boundaries [Fig.\ \ref{ES_EE2}(a)].
Thus, these boundary modes can give a subsystem-size dependent contribution
to the entanglement entropy.
If we naively subtract out the contribution from boundary modes
in the entanglement entropy, as shown in Fig.\ \ref{ES_EE2}(b),
the effective central charge we extract from the coefficient still depends on the total system size.

\begin{figure}[htbp]
\centering
\includegraphics[width= 0.6 \textwidth] {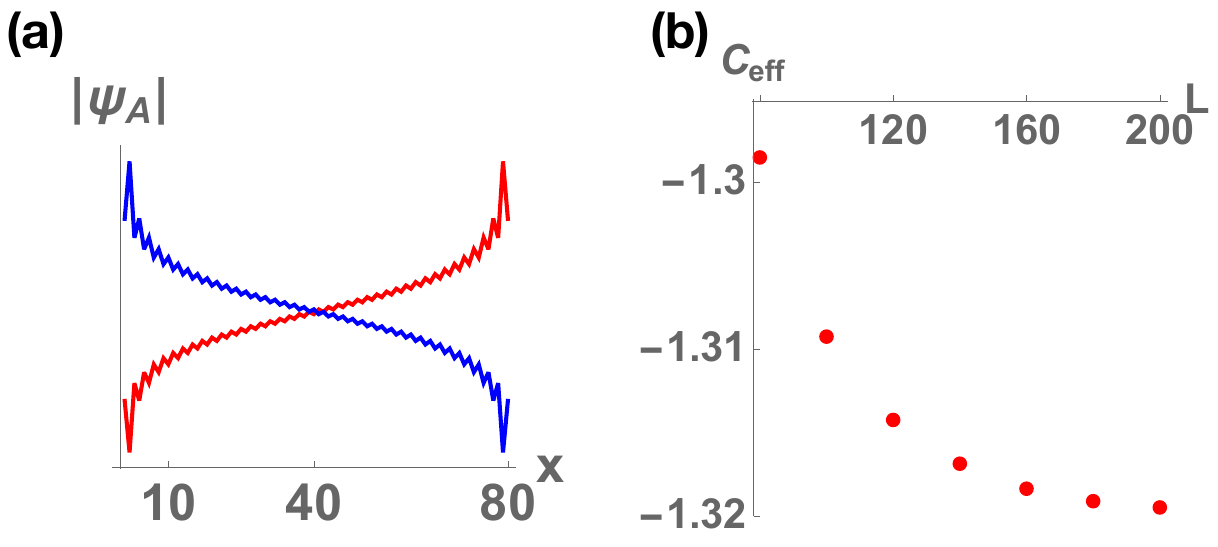}
\caption{(a) The boundary modes in the entanglement spectrum at the critical point $w-v=u$. (b) The effective central charge $c_{\rm eff}=3*\alpha$ after subtracting out the contributions of the boundary modes
as a function of total system size $L$. Here $\alpha$ being the coefficient in the
logarithmic scaling of the entanglement entropy, $S_A = \alpha \ln [\sin[\frac{\pi L_A}{L}]] + {\rm const.}$.
The effective central charge $c_{\rm eff}$ as a function of total system size $L$ for only including  the imaginary part of the  eigenvalues of the boundary modes in the entanglement spectrum. }
\label{ES_EE2}
\end{figure}

We observe the eigenvalues of the boundary modes
are $\xi_{\pm, \alpha}=0.5 \pm i  I_\alpha$ with $ I_\alpha$
depending on the subsystem size.
If we shift the unit-cell by half of the lattice constant,
the boundary modes in the entanglement spectrum can be removed
and the spectrum is identical to the critical point
that separates the trivial PT symmetric phase and the PT broken phase.
The entanglement entropy scaling gives the central charge $c=-2$. 

\section{Symmetry Enriched CFT}
In this section, we present the entanglement spectrum and entanglement entropy
analysis on the free-fermion version of the symmetry enriched CFT studied
in Ref.\ \cite{Verresen2019}.
The Hamiltonian for this critical chain [show in the left panel in Fig.\ \ref{CSPT}(a)] is 
\begin{align}
\mathcal{H}(k)=
\left(
\begin{array}{cc} 
0 & e^{i k} + e^{i 2 k} \\
e^{-i k} + e^{-i 2 k} & 0
\end{array}
\right).
\end{align}
One can immediately see
that there will be two isolated sites when the open boundary condition
is imposed.
It is shown that these boundary modes are exponentially localized
as long as the parity and time-reversal symmetries are preserved
in the bulk
\cite{Verresen2018}.
On the other hand, if we introduce boundaries by cutting through
the unit-cell,
or equivalently shifting the half of the unit-cell
of the original chain,
the system becomes the regular critical chain
[see right panel in Fig.\ \ref{CSPT}(a)].

We compute the entanglement spectrum and entropy in this symmetry-enriched CFT.
There are two mid-gap states which are localized at the boundaries of the
entanglement Hamiltonian [Fig.\ \ref{CSPT}(b)].
Due to these boundary modes,
 the entanglement entropy does not have logarithmic scaling and we cannot extract the central charge. 

 However, if we bipartite the system such that
 the entangling boundaries cut through the unit-cell,
 there is no boundary mode in the entanglement spectrum
 and the central charge $c=1$ can be directly extracted
 from the entanglement entropy scaling [Fig.\ \ref{CSPT}(c)]. 
\begin{figure}[htbp]
\centering
\includegraphics[width= 0.7 \textwidth] {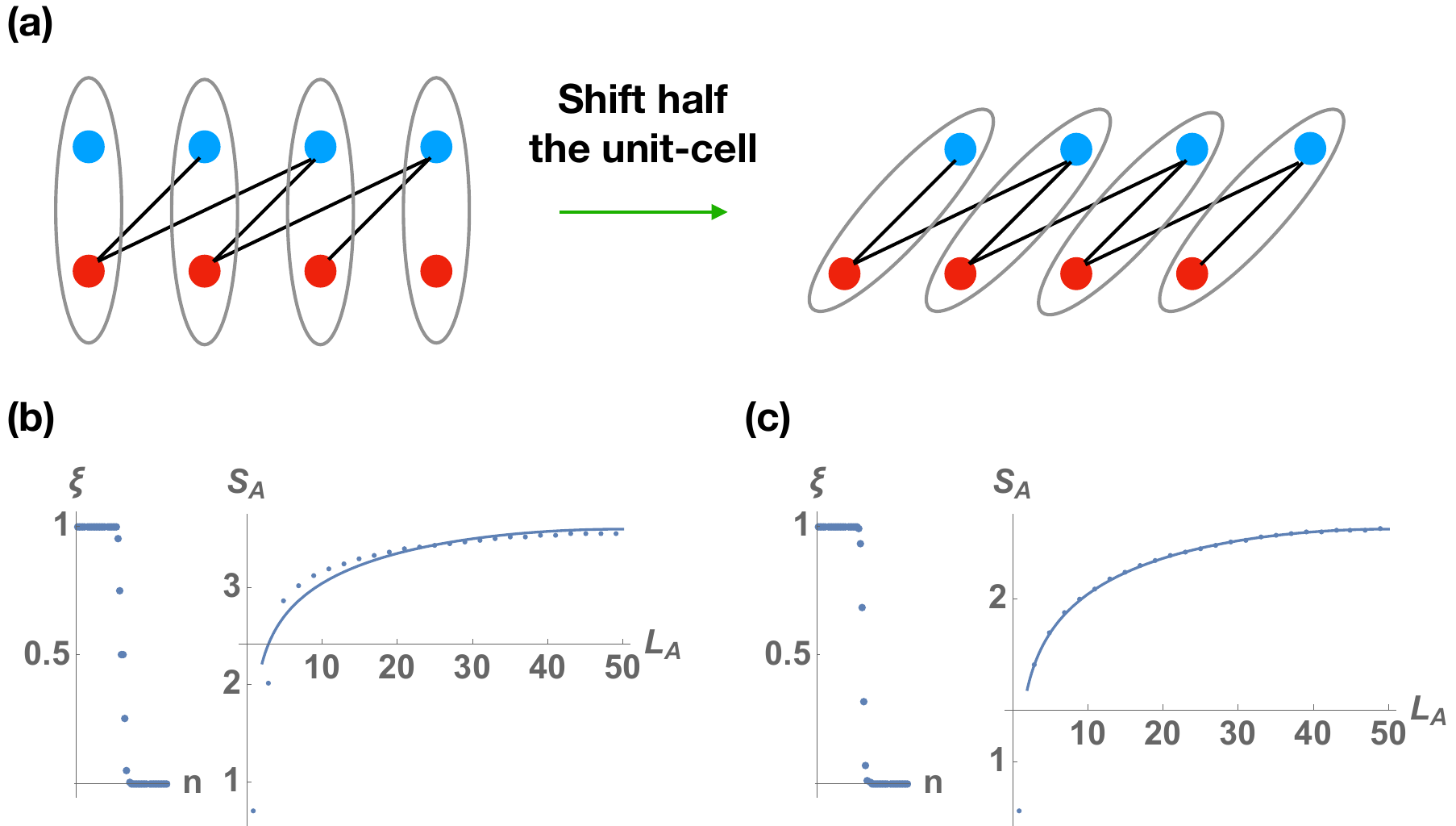}
\caption{(a) The free-fermion version of symmetry enriched CFT, which has  boundary modes in the critical chain as shown in left panel. After shifting half of the unit-cell, it has the regular critical chain configuration and has no boundary modes (right panel). (b) The entanglement spectrum (left) and the entanglement entropy scaling (right) in the symmetry enriched CFT. There are two mid-gap states in the entanglement spectrum corresponding to two boundary modes in the entanglement Hamiltonian. The entanglement entropy scaling does not satisfy the logarithmic scaling. (c) The entanglement spectrum (left) and the entanglement entropy scaling (right) in the critical chain case. There are no mid-gap states in the entanglement spectrum and the entanglement entropy scaling gives the central charge $c=1$. }
\label{CSPT}
\end{figure}

\section{{\it bc}-ghost CFT}
\label{Sec: bc_ghost CFT}
In this section, we briefly review the {\it bc}-ghost CFT
\cite{Polchinski,Blumenhagen,FRIEDAN198693, KAUSCH2000513, Kausch1995,GURUSWAMY1998661}.
The effective action for the {\it bc}-ghost theory is 
\begin{align}
S= \int d^2z  (\psi_b \bar{\partial} \psi_c+  \bar{\psi}_b \partial \bar{\psi}_c),
\end{align} 
where $\psi_{b/c}$ are the fermionic ghost fields for the right moving mode
and $\bar{\psi}_{b/c}$ denotes the anti-holomophic fermionic ghost fields which correspond to the left moving mode. 
We use the shorthand notations, $z= x + i t$, $\bar{z} = x - it$, 
$\partial =\frac{1}{2} (\partial_x- i\partial_t)$ and $\bar{\partial} = \frac{1}{2} (\partial_x+ i\partial_t)$.
Here we have the anti-commutation relationship $\{ \psi_b(z), \psi_c(w) \} = \delta(z-w)$.
From dimensional analysis, the conformal dimension of the ghost fields must be $\Delta_b+\Delta_c =1$.
We can parametrize them by $\Delta_b= \lambda$ and $\Delta_c= 1- \lambda$.
The equations of motion give
\begin{align}
&\bar{\partial}\psi_b(z) = \bar{\partial}\psi_c(z)=0, \notag\\
&\bar{\partial}\psi_c(z) \psi_b (0) =2\pi \delta^{(2)} (z,\bar{z}).
\end{align}
The above equations imply the operator product expansion (OPE) and the 
 two-point function  are 
 \begin{align}
 &\psi_b (z)  \psi_c (w)\sim  \frac{1}{z-w}, \notag\\
 &\langle \psi_b(z) \psi_c (w) \rangle \sim \frac{1}{z-w}.
 \end{align}
The other two-point functions do not have singularity, $\langle \psi_b (z) \psi_b (w) \rangle \sim \langle \psi_c (z) \psi_c (w) \rangle \sim O(|z-w|)$.

The Noether's theory gives the normal-ordered holomorphic part of the energy-momentum tensor,
\begin{align}
T(z) = : (\partial \psi_b) \psi_c: - \lambda  \partial :(\psi_b \psi_c):.
\end{align}
The OPEs between $T$ and $\psi_{b/c}$ are
\begin{align}
T(z) \psi_b(w) &\sim  (\partial \psi_b(z)) \psi_c(z)  \psi_b(w) - \lambda  \partial (\psi_b(z) \psi_c(z)\psi_b(w)) \notag\\
&\sim  \frac{1}{z-w}  \partial \psi_b(z) -\lambda\partial (\psi_b(z) \frac{1}{z-w}) \notag\\
&\sim \frac{\lambda}{(z-w)^2} \psi_b(w)+ \frac{1-\lambda}{z-w}\partial \psi_b(w),
                                                                                               \nonumber \\
T(z) \psi_c(w) &\sim  (1-\lambda )\partial (\psi_b(z) \psi_c(z)  \psi_c(w)) -   \psi_b(z) (\partial  \psi_c(z) )\psi_c(w) \notag\\
&\sim   (1-\lambda )\partial (\frac{-1}{z-w} \psi_c(z)) + \frac{1}{z-w} \partial  \psi_c(z) \notag\\
&\sim \frac{1-\lambda}{(z-w)^2} \psi_c(w)+ \frac{\lambda}{z-w}\partial \psi_c(w),
\end{align}
which give the conformal dimensions $\Delta_b =\lambda$,
and $\Delta_c =1-\lambda$, respectively.

The central charge can be obtained from the OPE of $T(z)T(w)$:
\begin{align}
T(z) T(w) &\sim  [:(\partial \psi_b) \psi_c: - \lambda  \partial :(\psi_b \psi_c):](z)  [:(\partial \psi_b) \psi_c: - \lambda  \partial :(\psi_b \psi_c):](w) \notag\\
&\sim  \frac{-1}{(z-w)^2}\partial_z \psi_b(z) \psi_c(w) + 6\lambda(1-\lambda)\frac{1}{(z-w)^4}  \notag\\
&\sim (-6\lambda^2 +6 \lambda -1) \frac{1}{(z-w)^4}.
\end{align}
We can identify the central charge $c = -12\lambda^2 +12 \lambda -2$.
In the non-Hermitian SSH model at the critical point, $(\Delta_b,\Delta_c)=(1, 0)$, which gives $c=-2$.

\section{Two-point functions}

Here, 
we compute the two-point functions in the non-Hermitian SSH model at the critical point $v-w=u$.
We first compute the correlation function $\langle  \psi^\dagger_b(x)  \psi_c(y) \ \rangle$ with $  \psi^\dagger_b(x) $ is the right creation operator
and $\psi_c(y) $ is the left annihilation operator. We refer these fields the ghost fields. 
\begin{align}
\langle  \psi^\dagger_b(x)  \psi_c(y) \ \rangle = {\rm Tr} \sum_{k} \frac{1}{L}e^{i k (x-y)} |R_{k,-}\rangle  \langle   L_{k,-}|
= \sum_{k} \frac{e^{i k (x-y)}}{L} = \frac{1}{\pi}\frac{\sin{\pi (x-y)}}{|x-y|}
\end{align}
This two-point function gives the correct conformal dimensions of the ghost fields, $\lambda_b+\lambda_c =1$ [Fig. \ref{2point}(a)].

We can also compute the other two-point functions $\langle  \psi^\dagger_b(x)  \psi_b(y) \ \rangle$ and  $\langle  \psi^\dagger_c(x)  \psi_c(y) \ \rangle$
\begin{align}
\langle  \psi^\dagger_b(x)  \psi_b(y) \ \rangle  =  {\rm Tr} \sum_{k} \frac{1}{L}e^{i k (x-y)} |R_{k,-}\rangle  \langle   R_{k,-}|
= \sum_{k} \frac{e^{i k (x-y)}}{L} ( \sin \frac{\phi_k}{2} \sin \frac{\phi^*_k}{2}+ \cos \frac{\phi_k}{2} \cos \frac{\phi^*_k}{2}), \notag\\
\langle  \psi^\dagger_c(x)  \psi_c(y) \ \rangle  =  {\rm Tr} \sum_{k} \frac{1}{L}e^{i k (x-y)} |L_{k,-}\rangle  \langle   L_{k,-}|
= \sum_{k} \frac{e^{i k (x-y)}}{L} ( \sin \frac{\phi_k}{2} \sin \frac{\phi^*_k}{2}+ \cos \frac{\phi_k}{2} \cos \frac{\phi^*_k}{2}),
\end{align}
where $\phi_k = \tan^{-1} [\frac{|w e^{-i k }+ v|}{i u}$]. As shown in Fig. \ref{2point}(b), there is no power law decay
as expected in the CFT.
\begin{figure}[htbp]
\centering
\includegraphics[width= 0.9 \textwidth] {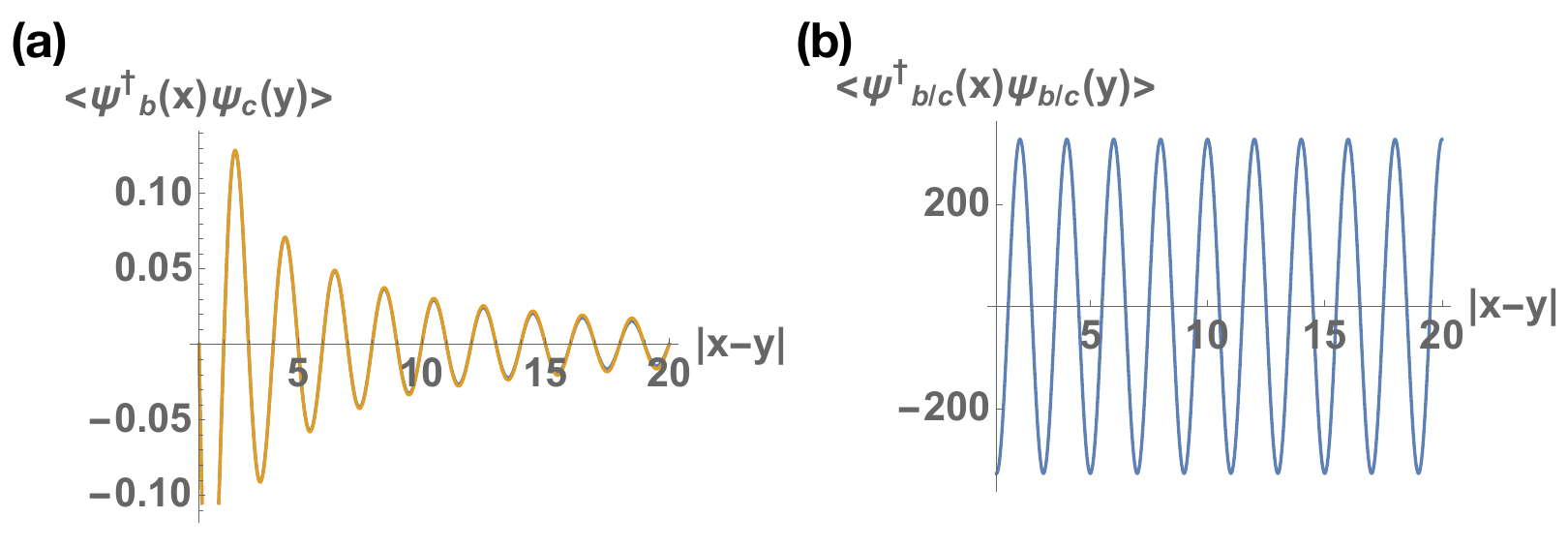}
\caption{ Two-point functions of (a) $\langle  \psi^\dagger_b(x)  \psi_c(y) \ \rangle$ and (b) $\langle  \psi^\dagger_{b/c}(x)  \psi_{b/c}(y) \ \rangle $
as a function of $|x-y|$.}
\label{2point}
\end{figure}

\section{Entanglement entropy in a (1+1)D non-unitary CFT}

The field theory approach to 
the entanglement entropy in non-unitary CFTs 
has previously been studied in, e.g., Refs.\ \onlinecite{Bianchini2015,Bianchini2015a,Narayan2016}, 
where the twist operator and replica method are used. 
In Ref.\ \onlinecite{Bianchini2015}, it was found that for a non-unitary CFT in which the physical ground state is different
from the conformal vacuum, the entanglement entropy has the form $S_A\sim c_{\text{eff}}\log\frac{l}{\epsilon}$, where $c_{\text{eff}}$ is the effective central charge, 
$l$ is the length of subsystem $A$, and $\epsilon$ is a UV cutoff.
In particular, it is found that $c_{\text{eff}}=c-24\Delta$, where 
$\Delta<0$ is the lowest conformal dimension of operator in the theory.
This result is a reminiscent of the work of by Itzykson, Saleur and Zuber \cite{Itzykson_1986},
where it was found that the central charge $c$ is replaced by the effective 
central charge $c_{\text{eff}}$ 
in the expression of the ground state free energy.
Later in Ref.\ \onlinecite{Narayan2016}, it was found that the entanglement entropy in the ghost 
$bc$ CFT with $c=-2$ has the form $S_A\sim c \log\frac{l}{\epsilon}$ with $c=-2$.
The underlying reason is that in this case the physical ground state is the same as the conformal vacuum (see more details in the following discussions). 

Here we give a brief review of these results on the entanglement entropy 
in non-unitary CFTs by utilizing the approach introduced by Cardy and Tonni
\cite{Cardy_2016}. 
Let us first introduce the possible difference between the physical ground state and the conformal vacuum.
In (1+1)D conformal field theory, the conformal vacuum $|0\rangle$ 
is defined as the
state invariant under all regular conformal transformations, $L_n|0\rangle=0$,
where  $n \ge -1$ (This results from the requirement that the stress-energy tensor $T(z)$ is regular at $z=0$ in the conformal vacuum). Therefore, for the conformal vacuum, we always have $L_0|0\rangle=0$.
On the other hand, the physical vacuum $|G\rangle$, 
or the physical ground state, 
is defined as the lowest eigenstate of $L_0$. In non-unitary CFTs, we have
\beq\label{GroundState}
L_0 |G\rangle= \Delta |G\rangle, \quad \overline{L}_0|G\rangle=\Delta |G\rangle,
\eeq
where $\Delta\le 0$ in general.
Here $L_0$ and $\overline{L}_0$ have the same lowest eigenvalue because 
the ground state is translation invariant.
In a unitary CFT, we always have $|G\rangle=|0\rangle$, which is not
necessarily true for a non-unitary CFT.

In the following derivation of the entanglement entropy, 
we will only use the conformal symmetry as well as the 
definition of physical ground state $|G\rangle$ of a generic non-unitary CFT
in Eq.\ \eqref{GroundState}.

For simplicity, we consider a finite interval $A=[-l/2,l/2]$ in an infinite system in the ground state $|G\rangle$.
The path-integral representation of the reduced density 
$\rho_A=\text{Tr}_B \rho$ 
can be expressed as
\cite{Cardy_2016}
\begin{eqnarray}\label{rho_A}
\small
\begin{tikzpicture}[baseline={(current bounding box.center)}]

\draw (30pt,0pt) arc (160:-160:3pt) ;
\draw (-30pt,0pt) arc (20:340:3pt) ;

\draw [thick][gray](-30pt,0.8pt)--(30pt,0.8pt);
\draw [thick][gray](-30pt,-2pt)--(30pt,-2pt);
\draw [dashed](-35pt,-1pt)--(-60pt,-1pt);
\draw [dashed]( 35pt,-1pt)--(60pt,-1pt);

\draw (75pt, 27pt)--(75pt,37pt);
\draw (75pt, 27pt)--(85pt,27pt);
\node at (81pt,32pt){$z$};

\node at (0pt,8pt){$A$};
\node at (50pt,8pt){$B$};
\node at (-50pt,8pt){$B$};

\node at (0pt,-20pt){$\,$};
\end{tikzpicture}
\end{eqnarray}
where $z=x+i\tau$, 
and two small discs of radius $\epsilon$ have been removed at the two
entanglement cuts $x=\pm R=\pm \frac{l}{2}$ as the UV cutoff. The
rows and columns of the reduced density matrix $\rho_A$
are labelled by the values of the fields on the upper and lower edges of the slit along $A$.
Along the two small discs, the conformal boundary conditions $|a\rangle$
and $|b\rangle$ are imposed.
Then, by considering the conformal transformation
$
w=f(z)=\log\frac{z+R}{R-z}, 
$
$\rho_A$ in Eq.\ \eqref{rho_A} is mapped to the following cylinder in $w$-plane:
\begin{eqnarray}\label{cylinder}
\begin{tikzpicture}[baseline={(current bounding box.center)}]

\draw [thick](0pt,0pt) ellipse (8pt and 20pt);
\draw [thick](80pt,0pt) ellipse (8pt and 20pt);
\draw (0pt,20pt)--(80pt,20pt);
\draw (0pt,-20pt)--(80pt,-20pt);

\draw [thick][gray](-8pt,1pt)--(72pt,1pt);
\draw [thick][gray](-8pt,-1pt)--(72pt,-1pt);

\draw [>=stealth,->] (105pt, 0pt)--(105pt,15pt);
\draw [>=stealth,->] (105pt, 0pt)--(120pt,0pt);
\node at (125pt,0pt){$u$};
\node at (110pt,15pt){$v$};

\draw (95pt, 27pt)--(95pt,37pt);
\draw (95pt, 27pt)--(105pt,27pt);
\node at (101pt,32pt){$w$};
\end{tikzpicture}
\end{eqnarray}
One can find that the length of the cylinder in $w$-plane is 
\beq
W=2\log\left(\frac{l}{\epsilon}\right)+\mathcal{O}(\epsilon),
\eeq
and the circumference is $2\pi$ in $v$ direction.
Then we have
$
\text{Tr}_A \rho_A^n=Z_n,
$
where $Z_n$ is the path integral over the manifold obtained by gluing $n$ cylinders in \eqref{cylinder} along the (gray) edges one by one. 
$Z_n$ can be explicitly evaluated as follows
\beq
Z_n=\langle a|e^{-H_{\text{CFT}}\cdot W}|b\rangle
=\langle a| e^{-\frac{2\pi}{2\pi n}(L_0+\bar{L}_0-\frac{c}{12})\cdot W}|b\rangle
=\sum_k\langle a|k\rangle \cdot \langle k|b\rangle \cdot e^{-\frac{1}{n}(\Delta_k+\bar{\Delta}_k-\frac{c}{12})\cdot W},
\eeq
where in the second step we have inserted a complete basis vectors $|k\rangle$.
Considering $W\gg 1$, only the lowest weight $\Delta_k+\bar{\Delta}_k$
dominate in $Z_n$.
Now the difference between unitary and non-unitary CFTs comes in. For a unitary CFT, the conformal vacuum is the same as the physical ground state, i.e.,
$|0\rangle=|G\rangle$. The term with $\Delta_k=\bar{\Delta}_k=0$ dominates, 
and therefore
\beq
Z_n\simeq \langle a|0\rangle \cdot \langle 0|b\rangle \cdot e^{\frac{c}{12 n}\cdot W},\quad \text{for unitary CFTs}
\eeq
On the other hand, for a non-unitary CFT, as seen in Eq.\ \eqref{GroundState}, the 
ground state has a possibly negative eigen-energy $\Delta\le 0$. Then $Z_n$ is dominated by the term with $\Delta_k=\bar{\Delta}_k=\Delta$:
\beq
Z_n\simeq 
\langle a|G\rangle \cdot \langle G|b\rangle \cdot e^{\frac{c-24\Delta}{12 n}\cdot W}
=:\langle a|G\rangle \cdot \langle G|b\rangle \cdot e^{\frac{c_{\text{eff}}}{12 n}\cdot W},\quad \text{for non-unitary CFTs}
\eeq
where we have defined 
\beq\label{c_eff}
c_{\text{eff}}=c-24\Delta.
\eeq
As a remark, the procedure of evaluating $Z_n$ here is essentially the same as 
the calculation of the free energy in the ground state of a non-unitary CFT as studied in Ref.\ \onlinecite{Itzykson_1986}.
Then the $n$-th Renyi entropy and von Neumann entropy can be expressed as
\beq
\begin{split}
S_{A}^{(n)}&:=\frac{1}{1-n}\log \frac{\text{Tr}(\rho_A^n)}{(\text{Tr}\rho_A)^n}
=\frac{1}{1-n}\log\frac{Z_n}{(Z_1)^n}
\simeq\frac{c_{\text{eff}}}{12}\cdot \frac{1+n}{n}\cdot W
\simeq\frac{c_{\text{eff}}}{6}\cdot \frac{1+n}{n}\log\left(
\frac{l}{\epsilon}
\right), \\
S_A&=\lim_{n\to 1}S_A^{(n)}\simeq
\frac{c_{\text{eff}}}{6}W
=\frac{c_{\text{eff}}}{3}\log\left(
\frac{l}{\epsilon}
\right),
\end{split}
\eeq
where we have neglected the $\mathcal{O}(1)$ terms which are contributed by the boundaries.
In particular, we have $c_{\text{eff}}=c$ for unitary CFTs, and $c_{\text{eff}}=c-24\Delta$ for non-unitary CFTs.

For the $bc$ ghost CFT with $c=-2$ as studied in the main text, it is noted that the physical ground 
state $|G\rangle$ is the same as the conformal vacuum $|0\rangle$, i.e., 
$
|G\rangle=|0\rangle$, and we have $\Delta=0$ in Eq.\ \eqref{GroundState}\cite{Narayan2016}.
Then based on Eq.\ \eqref{c_eff}, we have 
\beq
c_{\text{eff}}=c=-2.
\eeq
This agrees with the result obtained in Ref.\ \onlinecite{Narayan2016} based on the the twist
operator approach, with appropriate generalizations of the standard 
CFT replica technique.
Furthermore, for the ghost $bc$ CFTs with $\lambda>1$ (see previous sections), 
one has $\Delta=\frac{\lambda(1-\lambda)}{2}$ and the central charge
$c = -12\lambda^2 +12 \lambda -2$. The effective central
charge has the expression:
\beq
c_{\text{eff}}=c-24\cdot \frac{\lambda(1-\lambda)}{2},
\eeq
which reduces to $c_{\text{eff}}=c=-2$ for $\lambda=1$.

\section{Entanglement spectrum  of the non-Hermitian 2D model}

The non-Hermitian Chern insulator~\cite{Kawabata2018} defined by
the Bloch Hamiltonian 
\begin{align}
{\cal H}(\mathbf k)&=(m + t \cos k_x + t \cos k_y )\sigma_x  +(i \gamma + t \sin k_x )\sigma_y + (t \sin k_y )\sigma_z,
\end{align}
 has complex dispersion relations as
\begin{align}
E_{\pm}(\mathbf k) 
&= \pm
[
(m + t \cos k_x + t \cos k_y )^2 
+(i \gamma + t \sin k_x )^2 
+ (t \sin k_y )^2]^{1/2}.
\end{align}
Exceptional points appear when the two bands 
satisfy $E_{\pm}(\mathbf{k}_{\textmd{EP}})=0$. 
This condition demands $\sin (k_{\textmd{EP},x})= 0.$ We can find the gapless phases where pairs of exceptional points appear
on $k_x = 0$, $k_x = \pm \pi$, and both $k_x = 0$ and $k_x = \pm \pi.$

In the gapped phase with the topologically trivial bulk, no mid-gap modes appear between the gapped complex entanglement bands, as shown in Fig.~\ref{Fig_Trivial}.
In the gapless phases, there appear exceptional points on
$k_x = 0$ and/or $k_x = \pm \pi$. 
Complex entanglement spectra in the presence of exceptional points are shown in Fig.~\ref{Fig_ES2R} and Fig.~\ref{Fig_ES2R2L}.
\begin{figure}[htbp]
\centering
\includegraphics[width= 0.8 \textwidth] {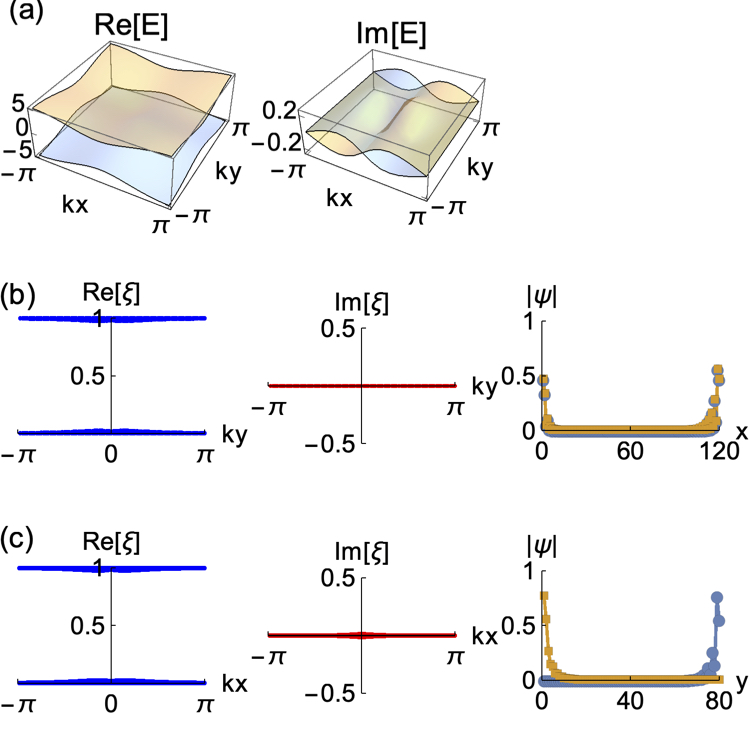}
\caption{The topologically trivial bands with zero Chern number ($t = 1.0, m = -3, \gamma = 0.5$). (a) The orange and blue bands
correspond to $E_{+}$ and $E_{-}$, relatively. (b) Complex entanglement spectrum which is labeled by $k_y$ and amplitude of the eigen-modes of $k_y=0$ for two degenerate $\xi= 0.981$~(yellow squares) and two degenerate $\xi=0.019$~(gray dots) along $x$ direction. (c) Complex entanglement spectrum labeled by $k_x$ and the amplitude of the eigen-modes of $k_x=0$ for  $\xi=0.0176 -0.013 i, 0.982 -0.013 i$~(yellow squares) and for $\xi=0.0176 +0.013 i, 0.982 +0.013 i$ (gray dots) along $y$ direction. }
\label{Fig_Trivial}
\end{figure}

\begin{figure}[htbp]
\centering
\includegraphics[width= 0.8 \textwidth] {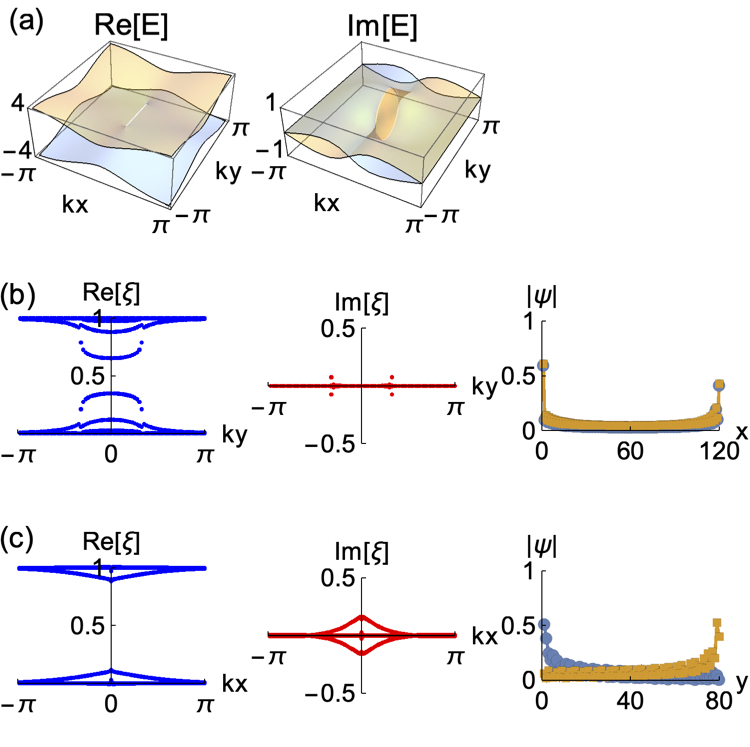}
\caption{The gapless phase with two pairs
of exceptional points on  both $k_x = 0$ ($t = 1.0, m = -2, \gamma = 1$). (a) The orange and blue bands
correspond to $E_{+}$ and $E_{-}$, relatively. (b) Complex entanglement spectrum which is labeled by $k_y$ and the eigen-modes of $k_y=0$ and  for $\xi=0.661 -0.001 i$~(yellow squares) and for $\xi= 0.339+ 0.001 i$ (gray dots) along $x$ direction. (c) Complex entanglement spectrum labeled by $k_x$ and the eigen-modes of $k_x=0$ for $\xi=0.147  +0.18 i$~(yellow squares) and for $0.853 -0.18 i$ (gray dots) along $y$ direction.  }
\label{Fig_ES2R}
\end{figure}

\begin{figure}[htbp]
\centering
\includegraphics[width= 0.8 \textwidth] {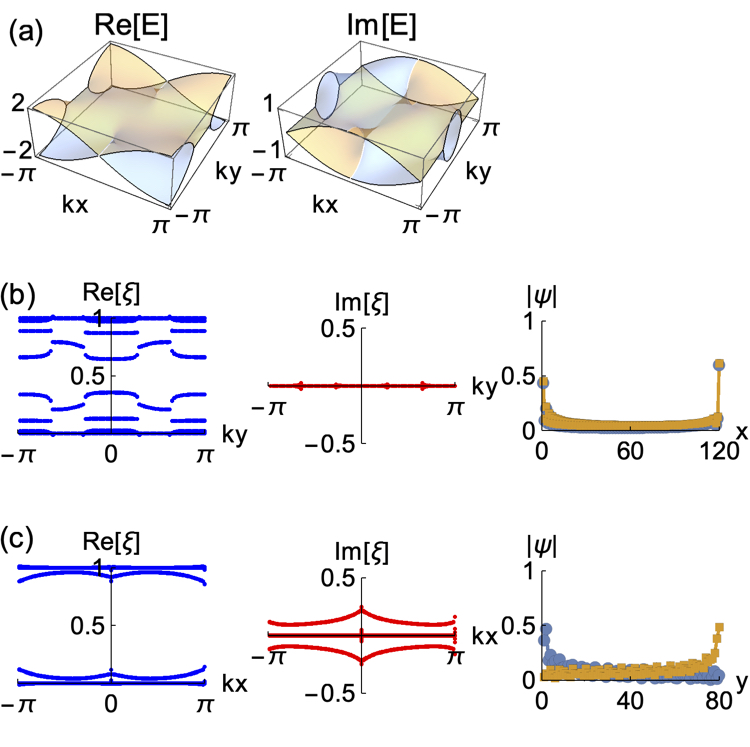}
\caption{The gapless phase with two pairs
of exceptional points on  both $k_x = 0$ and $k_x = \pm \pi$ ($t = 1.0, m = -0.2, \gamma = 1$). (a) The orange and blue bands
correspond to $E_{+}$ and $E_{-}$, relatively. (b) Complex entanglement spectrum which is labeled by $k_y$ and the eigen-modes of $k_y=0$ for $\xi=0.653 -0.002 i$~(yellow squares) and for  $\xi=0.347+ 0.002 i$~(gray dots) along $x$ direction. (c) Complex entanglement spectrum labeled by $k_x$ and the eigen-modes of $k_x=0$  for $\xi= 0.4+0.185i$~(yellow squares) and for $\xi=0.9 -0.185 i$~(gray dots) along $y$ direction. }
\label{Fig_ES2R2L}
\end{figure}
\end{widetext}

\end{document}